\documentclass{article}
\pdfoutput=1
\usepackage{cleveref}
\usepackage{epstopdf}
\usepackage{algorithm}
\usepackage{algpseudocode}
\usepackage{graphics}
\usepackage{epsfig}
\usepackage{subcaption}
\usepackage{booktabs}
\usepackage{tabularx}
\usepackage{amssymb}
\usepackage{mathtools}
\usepackage{colortbl}
\usepackage{collcell}
\usepackage{hhline}
\usepackage{pgf}
\usepackage{gensymb,siunitx}
\usepackage{etoolbox}
\usepackage{authblk}

\newcommand\gray{gray}

\newcommand\ColCell[1]{%
  \pgfmathparse{#1/100<.8?1:0}%
    \ifnum\pgfmathresult=0\relax\color{white}\fi
  \pgfmathparse{1-#1/100}%
  \expandafter\cellcolor\expandafter[%
    \expandafter\gray\expandafter]\expandafter{\pgfmathresult}#1}

\newcolumntype{E}{>{\collectcell\ColCell}c<{\endcollectcell}}

\newcommand{\name}{\texttt{Viden}}

\begin{document}
\title{Viden: Attacker Identification on \\In-Vehicle Networks} 

\author{Kyong-Tak Cho and Kang G. Shin}
\affil{University of Michigan, Ann Arbor\\
\{ktcho, kgshin\}@umich.edu}
\date{}                     
\setcounter{Maxaffil}{1}
\renewcommand\Affilfont{\itshape\small}


\maketitle
\begin{abstract}
	Various defense schemes --- which determine the presence of an 
	attack on the in-vehicle network --- have recently been proposed. 
	However, they fail to identify {\em which} Electronic Control Unit (ECU) 
	actually mounted the attack. 
	Clearly, pinpointing the attacker ECU is essential for fast/efficient 
	forensic, isolation, security patch, etc.
	To meet this need, we propose a novel scheme, called \name\ 
	(\underline{V}oltage-based attacker \underline{iden}tification), which can 
	identify the attacker ECU by measuring and utilizing voltages on the 
	in-vehicle 
	network. The first phase of \name, called {\em ACK learning}, determines 
	whether or not the measured voltage signals really originate from the 
	genuine message 
	transmitter. \name\ then exploits the voltage measurements to construct and 
	update the transmitter ECUs' voltage profiles as their fingerprints.
	It finally uses the voltage profiles to identify the attacker ECU. 
	Since \name\ adapts its profiles to changes inside/outside of the vehicle, 
	it can pinpoint 
	the attacker ECU under various conditions. Moreover, its efficiency and 
	design-compliance with 
	modern in-vehicle network implementations make \name\ practical and easily 
	deployable.
	Our extensive experimental evaluations on both a CAN bus prototype and two 
	real vehicles 
	have shown that \name\ can accurately fingerprint ECUs based solely on 
	voltage measurements 
	and thus identify the attacker ECU with a low false identification rate of 
	0.2\%.
\end{abstract}

\section{Introduction}
Remote and/or driverless control of a car is no longer science fiction.
In fact, demonstration and deployment of such a vehicle control have become 
prevalent,
triggering significant R\&D efforts and investments from industry, governments, 
and academia.
Despite their numerous benefits, these technological developments have created
serious safety/security concerns.

These concerns are genuine and real. For example, researchers
evaluated various remote access points on vehicles and
demonstrated that an attacker can exploit them to remotely compromise
Electronic Control Units (ECUs)~\cite{checkoway}.
By exploiting the compromised ECUs, researchers have shown that it is feasible 
to remotely control or even shut down a vehicle \cite{koscher, woot_tele, 
	checkoway, miller, busoff, jeepkill, teslahack}.

Numerous schemes have been proposed to detect and/or prevent various vehicle 
cyber
attacks~\cite{hoppe, entropy, canauth, sensanom, delaymac, mult, cids}.
Although these countermeasures are capable of determining whether or not
there is an intrusion in the in-vehicle network, they cannot determine {\em 
	which} ECU is actually mounting the attack, i.e., incapable of {\em 
	attacker 
	identification}. This is because in-vehicle networks are mostly configured 
	as
broadcast buses and their messages lack information on the transmitters.
An accurate attacker identification, however, is imperative as it provides a 
swift pathway for
forensic, isolation, security patch, etc.
No matter how well an Intrusion Detection System (IDS) detects the presence of 
an intrusion
in a vehicle, if we still do not know which ECU is mounting the attack and hence
which ECU to isolate/patch, the vehicle remains insecure and unsafe.
It is much better and more economical to isolate/patch the attacker ECU, than
blindly treating {\em all} ECUs as (possible) attackers.


To meet this essential need for attacker identification
--- that existing solutions have not yet been able to satisfactorily meet ---
we propose a novel scheme, called {\name} (\underline{V}oltage-based attacker 
\underline{iden}tification), which fingerprints message 
transmitter ECUs on Controller Area Network (CAN) via voltage measurements and 
thus facilitates attacker identification.
Of the various in-vehicle network protocols, we focus on CAN as it is the {\it 
de facto} standard
for in-vehicle networks and its adoption has been mandated in all cars 
manufactured since 2008~\cite{canmandatory}.
The rationale behind using voltage for fingerprinting ECUs is the existence of 
small inherent
discrepancies in different ECUs' voltage outputs when they inject messages.
To capture this and then use it to fingerprint the transmitter ECUs, \name\ 
first monitors the output
voltages from the two dedicated wires on the CAN bus: CAN-High (CANH) and 
CAN-Low (CANL).
All ECUs' transceivers are connected to, and use these for their message 
transmissions and receptions. Through the acquired voltage measurements for 
each message ID, \name\ first learns the {\em ACK threshold}, the key 
information \name\ uses to discard the measurements of voltages outputted by 
ECUs 
while acknowledging the receipt of, but not transmitting, the message.
\name\ utilizes the thus-derived ACK threshold to learn the voltage output 
behavior of each in-vehicle ECU 
by constructing new features called {\em voltage instances}. Then, it 
transforms those
instances to the transmitter ECU's voltage profile (i.e., {\em fingerprint}) 
via Recursive Least Square (RLS) algorithm, an adaptive signal 
processing technique.
As a result, \name\ utilizes the derived voltage profiles for an accurate 
attacker identification.
Through experimental evaluations on a CAN bus prototype and on two real 
vehicles,
we show that the constructed voltage profiles are distinct for different ECUs, 
thus validating
\name's capability of identifying the attacker ECU.

While there have been proposals to fingerprint ECUs with timing~\cite{cids} 
or voltage (like \name) measurements~\cite{sourceiden, voltarxiv}, their 
practicality and efficiency in identifying the attacker ECU remain limited to 
only certain attack scenarios, mainly because they were designed for intrusion 
detection, not attacker identification. In other words, there are many 
scenarios 
in which existence of an attack is detected but the 
attacker cannot be identified correctly.
Thus, we design, implement, and evaluate \name\ by focusing on attacker 
identification via a distinct way of fingerprinting ECUs from the existing 
schemes. 
As a result, \name\ is efficient and easy to deploy on any ECU, thanks 
to its {\em adaptability} and {\em practicality}.
\vspace{-0.5em}
\paragraph*{Adaptability.}
Existing voltage-based fingerprinting uses supervised
batch learning that generates a norm model by learning from a pre-defined
training data set~\cite{sourceiden, voltarxiv}. So, until the training data set 
and hence
models/fingerprints are updated again, the norm models remain unchanged.
Such an approach, however, cannot adapt the norm models to unexpected changes
(e.g., changes in temperature) inside/outside the vehicle.
More importantly, adversaries who intentionally generate changes
can evade these existing fingerprinting schemes.
\name\ takes a very different approach from them in that it models and 
updates the voltage-based fingerprints by applying adaptive signal processing 
(i.e., {\em online} (not batch) learning) to its new set of features: voltage 
instances.
This enables \name\ to correctly modify the fingerprints and hence adapt to
inevitable but unpredictable changes in vehicles that can either occur
naturally (due to the mother nature) or be intentionally triggered by an 
intelligent adversary.
Such adaptability is essential for vehicle security.

\vspace{-0.7em}
\paragraph*{Practicality.}
Unlike the existing voltage-based fingerprinting schemes, the unique approach 
taken by \name\ 
eliminates the requirement/assumption of using a 
specific CAN message type or CAN bus speed, thus facilitating its 
deployment.
Moreover, it does not require any knowledge of which message fields the 
voltages are measured on, i.e., {\em message-field-agnostic}.
This enables \name\ to achieve its goal even with a low voltage sampling rate, 
thus lowering cost.
Furthermore, even though it is message-field-agnostic, since \name\ filters
out undesired samples using its derived ACK thresholds, there is no need
to impose restrictions on which fields of the message should be sampled to run 
\name. All of these salient features enable \name\ to run without re-designing 
current
CAN controllers and make \name\ very practical and cost-efficient, which 
is very important for the cost-conscious automotive industry.

We have implemented and evaluated \name\ on a CAN bus prototype and on two real 
vehicles. Our evaluation results show that \name\ can identify the attacker ECU 
with a low false identification rate of 0.2\%, thanks to its unique 
fingerprinting that makes it adaptive to handle various attack scenarios.

This paper makes the following main contributions:
\begin{enumerate}
	\item Proposal of a new scheme which retains only the voltage measurements 
	output by the transmitter ECU (Section~\ref{sec:learnack});
	\item Design of \name\ which constructs voltage profiles, i.e., 
	fingerprints, 
	by 
	modeling the norm voltage output behaviors of in-vehicle ECUs and exploits 
	them 
	for accurate identification of the attacker ECU
	(Sections~\ref{sec:phase2}--\ref{sec:adjust}); 
	\item Implementation and demonstration of \name\ on a CAN bus prototype and 
	on 
	two real vehicles (Section~\ref{sec:evaluation}).
\end{enumerate}


\section{Background}\label{sec:primecan}
\subsection{CAN Message Transmission}
\begin{figure}[t]
	\centering
	\begin{subfigure}[t]{\linewidth}
		\includegraphics[width=\linewidth]{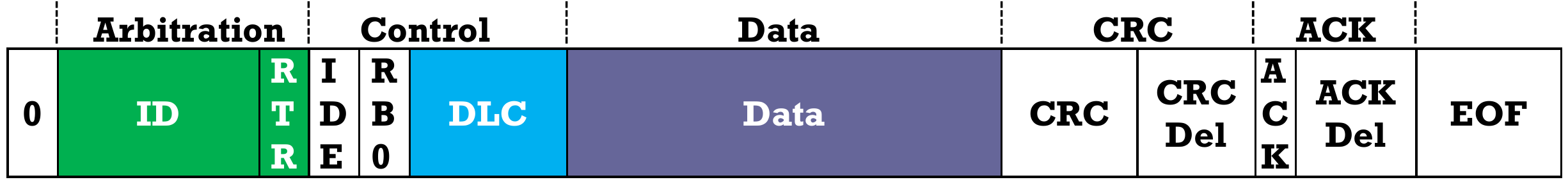}
		\caption{Format of a standard CAN data frame.}
		\label{fig:canframe}
	\end{subfigure}%
	\\
	\begin{subfigure}[t]{\linewidth}
		\centering
		\includegraphics[width=0.8\linewidth]{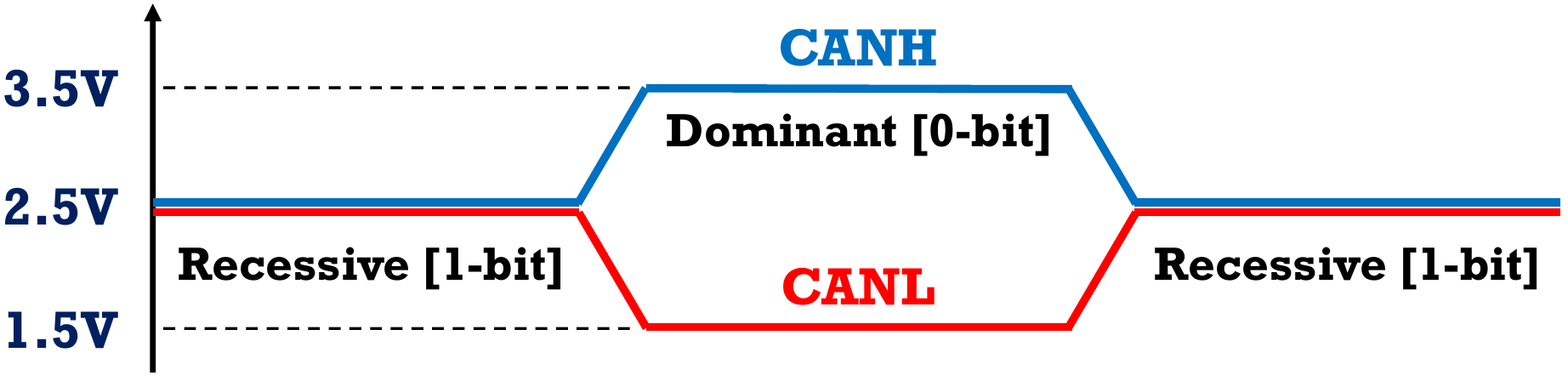}
		\caption{CAN output voltages when sending a message.}
		\label{fig:canvolt}
	\end{subfigure}%
	\caption{Message transmission via outputting voltages.}
\end{figure}

In-vehicle ECUs broadcast their retrieved sensor data 
via a CAN frame/message. Instead of carrying the address of the 
transmitter/receiver, as shown in Fig.~\ref{fig:canframe}, it contains a unique 
identifier (ID), which represents its priority. 
Starting from a 0-bit followed by a sequence of dominant (0) or recessive 
(1) bits, all fields within the CAN frame are sent on the bus by the 
``transmitter ECU'' except for the Acknowledgment (ACK) slot.
The ACK slot is, in fact, used by {\em all} ECUs {\em at the same time} ---
except for the transmitter ECU --- that have correctly received the preceded 
fields of the ACK slot, regardless of whether they are 
interested in their content or not. If correctly received, those ECUs send a 
0-bit in the ACK slot. 
Thus, multiple ECUs acknowledge the message simultaneously, even before 
the transmitter finishes sending its message on the bus. 

To send either a 0- or 1-bit, CAN transceivers (are agreed to) output 
certain voltage levels on the two dedicated CAN wires: CANH and CANL.
As shown in Fig.~\ref{fig:canvolt}, to issue a 0-bit on the 
CAN bus, CAN transceivers (are agreed to) output approximately 3.5V on CANH
and 1.5V on CANL so that the differential voltage becomes approximately 2V.
On the other hand, when sending a 1-bit, the transceivers output
approximately 2.5V on both CANH and CANL, yielding a differential 
voltage of approximately 0V~\cite{an228, iso2}.
So, by measuring the differential voltage of CANH and CANL,
receiver ECUs read the streams of 0 and 1 bits, and thus receive the message.
From this perspective, CAN is a {\em differential bus}.

\begin{figure}[t!]	
	\centering
	\begin{subfigure}[t]{0.47\linewidth}
		\centering
		\includegraphics[height=5.5cm]{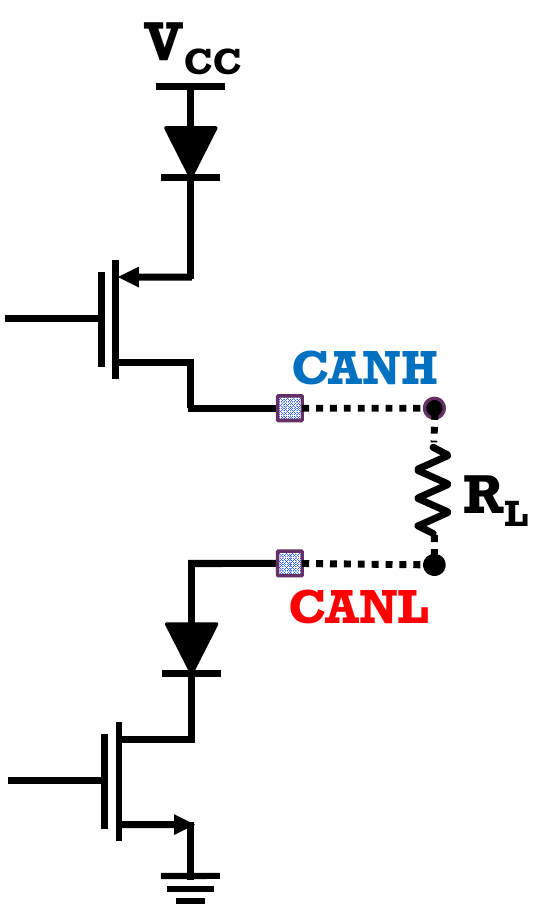}
		\caption{Transceiver schematic.}
		\label{fig:cantrans1}
	\end{subfigure}%
	\hspace{0.3cm}
	\begin{subfigure}[t]{0.47\linewidth}
		\centering
		\includegraphics[height=5.5cm]{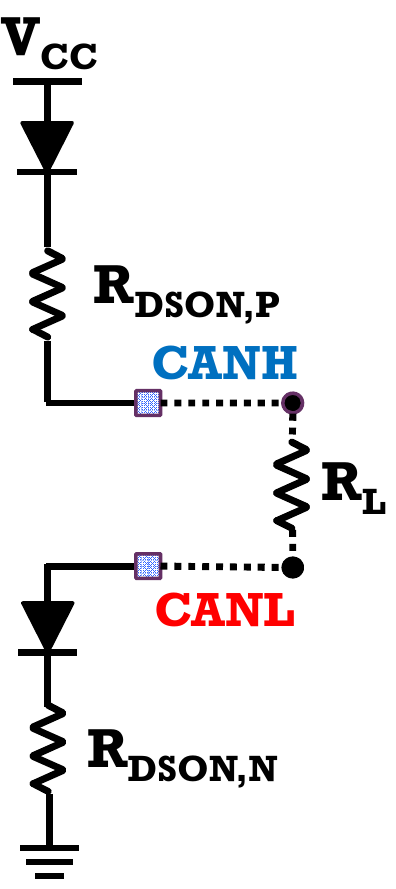}
		\caption{When sending a 0-bit.}
		\label{fig:cantrans2}
	\end{subfigure}
	\caption{Output schematics of a CAN transceiver.}
\end{figure}
%
CAN transceivers output the intended voltages by simultaneously 
switching on/off their transistors.
Fig.~\ref{fig:cantrans1} shows an equivalent schematic of a CAN 
transceiver~\cite{ti1,ti2}.
Note that CAN transceivers of multiple ECUs are connected 
to the CAN bus in parallel, thus sharing the same load resistance $R_L$, which 
is normally 
set to $60\Omega$~\cite{vector}.
The high- and low-side output circuits consist of a series diode and a P- and 
N-channel transistor, respectively.

For the transceiver to send a 1-bit, both the high and low side transistors are 
switched {\em off} and are thus in a high impedance state. This results in 
negligible current flowing from $V_{CC}$ to ground, yielding negligible 
differential voltage on CANH and CANL.
On the other hand, when sending a 0-bit, both transistors are turned {\em on} 
and are thus in a low impedance state. When the transistors are on, they can be 
equivalently described as resistors with drain-to-source on-state 
resistance $R_{DSON}$ as shown in Fig.~\ref{fig:cantrans2},
where current flows from $V_{CC}$ to ground through $R_L$ and thus creates a 
differential voltage of (approximately) 2V between CANH and CANL. 
This way, the CAN transceivers are capable of outputting either 0 or 2V of 
differential voltage on the two CAN wires.

\subsection{Related Work}\label{sec:relworks}
Researchers have attempted to fingerprint ECUs in various ways, mostly for the 
purpose of intrusion detection.

A clock-based intrusion detection system (CIDS) was proposed in \cite{cids} to
detect intrusions by fingerprinting ECUs on CAN. 
CIDS derived the fingerprints by extracting the ECUs' {\em clock skews} from 
message 
arrival times.
While the main objective of CIDS was to detect intrusions, the authors of 
\cite{cids} mentioned that the thus-derived fingerprints may also be used for 
attacker identification, but only when attack messages are injected {\em 
	periodically}. In other words, if the attacker transmits messages {\em 
	aperiodically}, then CIDS cannot identify the attacker ECU, i.e., the 
	adversary 
can evade CIDS as far as attacker identification is concerned.
\name\ takes an entirely different approach: looking at attack messages from 
the perspective of ECUs' {\em output voltages} on CAN. This allows \name\ to 
accurately identify 
the attacker ECU irrespective of how and when the attacker injects its 
messages, which is crucial for attacker identification.

Instead of fingerprinting ECUs based on message timings, as in \name, some 
researchers also proposed to fingerprint them via voltage measurements. 
The authors of \cite{sourceiden} used the Mean Squared Error (MSE) of 
voltage measurements as fingerprints of ECUs.
However, they were shown to be valid only for the voltages measured during the 
transmission of CAN message IDs, and more importantly when voltages were 
measured on a low-speed (10Kbps) CAN bus; 
this is far from contemporary vehicles that usually operate on a 500Kbps CAN 
bus. 

To overcome these difficulties, researchers proposed to extract other time and 
frequency domain
features of voltage measurements (e.g., RMS amplitude) and use them as
inputs for classification; more specifically, supervised (batch) learning 
algorithms (e.g., SVM) \cite{voltarxiv}.
This way, they were able to fingerprint ECUs with enhanced accuracy and was 
successful on
high-speed CAN buses. 
However, this solution was neither practical nor attractive for attacker 
identification for the following reasons.
First, it required not only a high sampling rate (2.5 GSamples/sec),
but also the use of the {\em extended} CAN frame format with 29-bit IDs, which 
is seldom used (due to its bandwidth waste) in contemporary vehicles; 
most vehicles use the {\em standard} format with 11-bit IDs.
Moreover, since the modeling was done via batch learning, unpredictable changes
in the CAN bus (e.g., temperature, battery level) and adversary's behaviors can 
lead to false identifications. These will be detailed later when we discuss the 
details of \name.

In contrast, \name\ fingerprints ECUs very differently and hence achieves
effective attacker identification
(1) through online update of fingerprints via adaptive signal processing to 
provide adaptability; 
(2) at a low sampling rate (50 KSamples/sec); and more importantly, (3) without 
imposing restrictions on the type of CAN message or the speed of CAN bus to be 
used. 
As a result, the deployment of \name\ in legacy and new vehicles will be much 
easier.
\section{Viden}\label{sec:volfing}
Attacker identification is essential for expedited forensic,
isolation, and security patches, all of which are the key requirements for
vehicle safety. To meet this need, we propose a novel fingerprinting scheme,
\name, that exploits small inherent discrepancies in different ECUs' voltage
outputs. Before delving into the inner workings of \name, we first describe the
system and threat models.

\subsection{System and Threat Models}\label{sec:attackmodel}
\subsubsection{System Model}
The vehicle's CAN bus under consideration is assumed to have been equipped with 
an IDS as well as 
a timing- (e.g., CIDS \cite{cids}) and voltage-based (e.g., schemes in 
\cite{voltarxiv, sourceiden} 
or \name) fingerprinting device;  the latter complements the former via 
attacker identification.
We discern a fingerprinting device from an IDS based on the fact that the IDS
detects the {\em presence} of an attack whereas the fingerprinting device
identifies the {\em source} of the (detected) attack.
An attack can be mounted by the adversary who has control of a
physically/remotely compromised ECU. In our system model, however,
we consider such an ECU to have been {\em remotely} compromised and
thus controlled by the adversary as in \cite{checkoway, miller3}. We do not
consider a compromised device which was attached to the in-vehicle
network (e.g., device plugged in the OBD-II port), as it requires physical 
access and
its identification has been addressed elsewhere \cite{busoff2,voltarxiv}.
So, the compromised ECU we consider is one of those originally
installed on the vehicle's CAN bus.

\subsubsection{Threat Model}
By injecting fabricated attack messages through his compromised ECU, the
attacker can control the vehicle maneuver.
We consider the attacker to be smarter than this: beyond just controlling the 
vehicle, the
attacker's goal is to also hide the identity of the ECU injecting the attack 
messages. 
That is, while the deployed IDS may detect the presence of an attack,
the adversary tries to {\em evade} the fingerprinting device, i.e., prevent it
from determining the source of the attack.
For evasion, the adversary can perform two different impersonations when
injecting his attack messages:
\begin{itemize}
	\item {\em Arbitrary impersonation}: The attacker misleads the
	fingerprinting device to think that some {\em arbitrary} ECU other than
	himself is the attacker.
	\item {\em Targeted impersonation}: The attacker acts
	smarter by impersonating a {\em targeted} ECU for evasion, i.e.,
	make the fingerprinting device believe that the targeted ECU is the 
	attacker.
\end{itemize}

Depending on the adversary's capabilities and knowledge of different
defense schemes (available in the market or in literature) as well as their
operation, his approach to evading a fingerprinting device would be different.
Specifically, based on whether the adversary is aware of the fact that an
in-vehicle ECU can be fingerprinted via timing and/or voltage measurements, his 
best
effort in achieving his goal would be different. Thus, we consider three
different types of adversaries: {\em naive}, {\em timing-aware}, and {\em
	timing-voltage-aware} adversaries.

While all attackers are capable of injecting and sniffing messages on the
CAN bus, a {\em naive} adversary does not have any knowledge of how ECUs can be
fingerprinted (either via timing or voltage), due possibly to lack of his
technical expertise or curiosity. Thus, the naive adversary injects his attack
messages imprudently at arbitrary times with forged message IDs (for
impersonation).

An intelligent adversary, however, might know how ECUs can be
fingerprinted via timing analysis.
Thus, the adversary uses his knowledge to evade any
(possibly-installed) fingerprinting scheme as much as possible as follows.
The adversary logs CAN traffic, learns the timing behavior of other ECUs,
and exploits the learned information in injecting attack messages at the
appropriate (learned) times so as to imitate other ECUs' timing behavior.
This way, the adversary can perform arbitrary/targeted impersonation and thus
attempt to evade the fingerprinting device. We refer to this adversary as a
{\em timing-aware} adversary.

The adversary might also have knowledge of how ECUs can be fingerprinted
via voltage and timing measurements.
Hence, when injecting attack messages, such an adversary may try to exploit his
knowledge in impersonating other ECU(s) and thus evade any
fingerprinting device as much as possible. We call this adversary a {\em
	timing-voltage-aware} adversary.
We consider such an adversary to be capable of changing his voltage outputs via
running battery draining processes, changing the supply voltage
level, or by heating up or cooling down the ECU.
Although he can change them to a certain level, we consider him to be
incapable of {\em precisely} controlling their instantaneous values. This
is reasonable as precise control of voltages would require, for
example, control of even the ambient temperature.
By changing his ECU's voltage outputs to a certain level in which the targeted 
ECU is 
outputting, a timing-voltage-aware adversary can perform a targeted 
impersonation.
Similarly, he can arbitrarily change the output levels for arbitrary 
impersonation.
Since changing his voltage levels (either before or during message
injections) does not necessarily imply that he is attacking the CAN bus, in
this paper, we differentiate ``impersonation'' from an actual attack of message 
injections.
In addition, the adversary might even know when the voltage-based fingerprints
are updated (if not updated in real time) and thus use that as a reference in
determining when to perform arbitrary/targeted impersonation.
Note, however, that he must "play" within the setting boundaries of the given
CAN bus. For example, the attacker cannot control/tune the values of resistors
within the CAN bus in order to control the voltage levels, as this requires 
physical access.

\begin{figure}[!t]
	\centering
	\includegraphics[width=\linewidth]{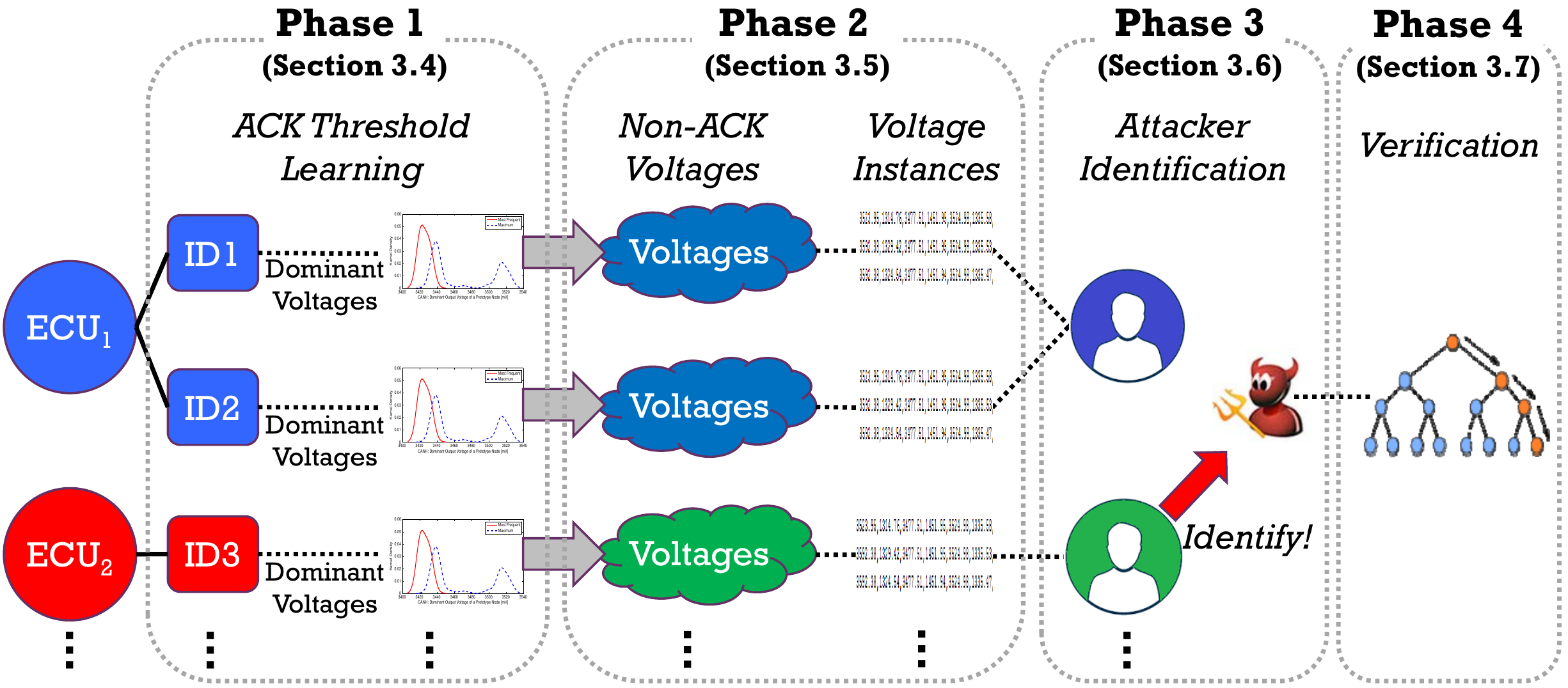}
	\caption{An overview of \name.}
	\label{fig:overview}
\end{figure}

\subsection{High-Level Overview of \name}
As shown in Fig.~\ref{fig:overview}, \name\ fingerprints ECUs via voltage
measurements and achieves attacker identification in four phases.

{\bf Phase 1:}
\name\ measures the CANH \& CANL voltages and maps the recently acquired
values to the ID of the message it has just received through the ECU's receive
buffer. Then, for that message ID, \name\ learns its ACK threshold. This
threshold helps \name\ determine whether or not the measured voltage
originates from the actual message transmitter. Phase 1 is run in the
initialization step of \name\ and when an update is necessary.

{\bf Phase 2:}
Exploiting the learned ACK threshold, \name\ selects voltages
that are outputted solely by the message transmitter. Then,  \name\ uses them
to derive a {\em voltage instance}, which is a set of features that reflect the
transmitter ECU's voltage output behavior. Phase 2 and onwards are run 
iteratively.

{\bf Phase 3:}
\name\ uses every newly derived voltage instance to update the voltage
profile of the message transmitter. When an attack is detected by the IDS,
\name\ constructs a voltage profile for the attack messages and maps that
profile to one of those \name\ has, thus identifying the attacker ECU.

{\bf Phase 4:}
The results from Phase 3 are verified further via multi-class classification,
only when necessary.

For a given message ID, only one ECU is assigned for its transmission in most
cases. Thus, for now we consider the relationship between the numbers of ECUs,
IDs, and voltage profiles to be 1, $N (\ge 1)$, and 1, respectively.
We will discuss further in Section~\ref{sec:disc} on how \name\ deals with
cases where the relationship between the numbers of  ECUs and IDs might be $N$
and 1, respectively.

\begin{figure}[t]
	\centering
	\includegraphics[width=\linewidth]{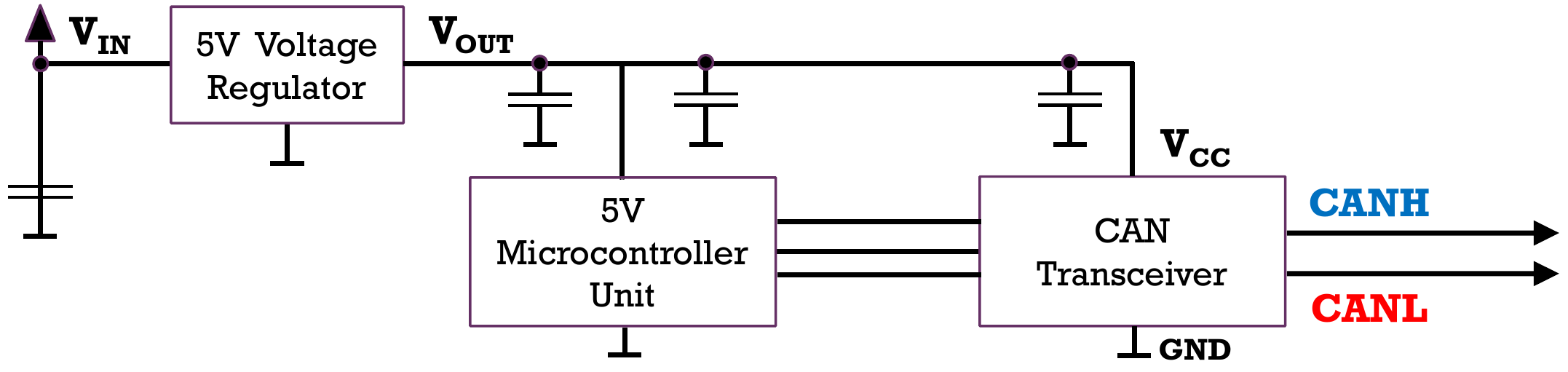}
	\caption{CAN typical application schematic.}
	\label{fig:canblock}
\end{figure}

\subsection{CANH and CANL Voltage Outputs}\label{sec:facts}
Before presenting the details of \name, we first discuss
which voltage characteristics of ECUs it exploits for attacker identification.

{\bf Variations in supply and ground voltages.}
Fig.~\ref{fig:canblock} shows a typical ECU connection to CAN~\cite{ti1,ti2}.
In order to output the desired voltage levels on CANH and CANL,
transceivers are powered with the nominal supply voltage ($V_{CC}$)
of 5V, which is provided and maintained by a voltage regulator.
The input of the regulator, $V_{IN}$, comes from a power supply, i.e.,
a 12V/24V battery powering all the ECUs~\cite{battery}.
Not only the voltage regulator but also the connected bypass capacitors help
stabilize the $V_{CC}$ level.
However, the output voltage of an ECU's regulator varies {\em independently}
and {\em differently} from other ECUs' regulators, as their supply
characteristics are different (e.g., different regulators' common-mode
rejection ratios).
Thus, there are inherent, small but non-negligible differences in ECUs'
$V_{CC}$. There exist variations in not only $V_{CC}$ but also in the ground
voltage since there does not exist a perfect ground~\cite{vector}.

For these reasons, CAN transceivers are built to operate over a range of
voltages (e.g., TI TCAN10xx devices are designed to handle 10\% supply 
variations~\cite{ti2}).
This guarantees transceivers with different $V_{CC}$ and/or ground to
communicate messages correctly.
\begin{figure}[!t]
	\centering
	\includegraphics[width=0.8\linewidth]{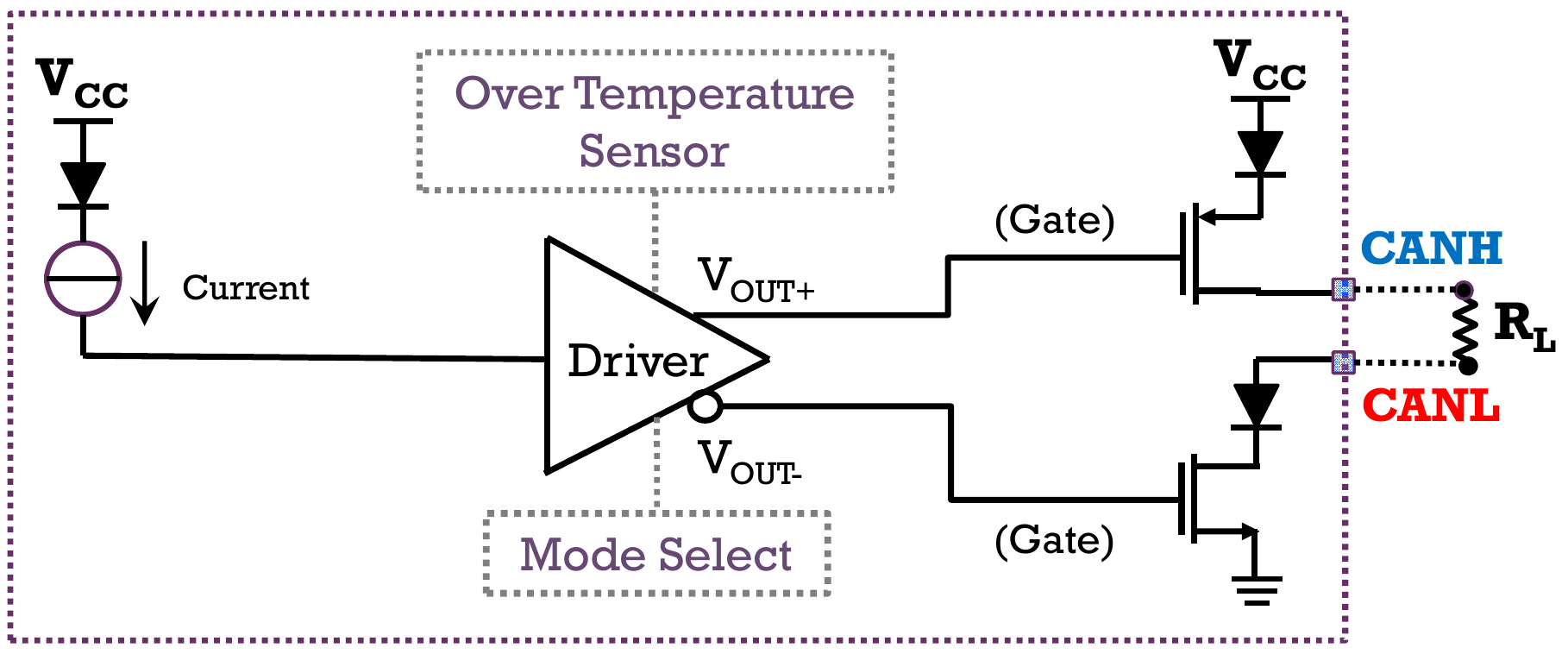}
	\caption{Transistors' gate voltages are fed by the driver.}
	\label{fig:canblock2}
\end{figure}

{\bf Variations in on-state resistance.}
When transceivers send a 0-bit, their two transistors are turned on so
that the flowing current generates the required
differential voltage between CANH and CANL.
In such a case, transistors in the transceivers are considered as resistors
$R_{DSON,P/N}$ (see Fig.~\ref{fig:cantrans2}).
Although transceivers are designed to have the same $R_{DSON,P/N}$
values, process/manufacturing variations/imperfections cause transistors'
$R_{DSON,P/N}$ values to be slightly different from each 
other~\cite{processvar}.

Fig.~\ref{fig:canblock2} shows a typical circuit diagram of a CAN transceiver.
Transistors' $R_{DSON}$ values are inversely related to their gate voltages,
which are supplied by a driver, i.e., a fully differential 
amplifier~\cite{gatevolt, mosfet}.
Interestingly, since the driver input is affected by $V_{CC}$, which also
varies with ECU, the transistors' gate voltages are also affected by $V_{CC}$.
Therefore, variations in $V_{CC}$ lead to variations in transistors'
actual $R_{DSON}$ values. In summary,
\begin{itemize}
	\item[$\mathbb{V}1$.] \emph{There exist differences/variations in CAN
		transceivers' nominal supply voltage, ground voltage, and $R_{DSON,P/N}$
		values, especially during the transmission of a 0-bit.}
\end{itemize}

When transmitting a 1-bit, the two transistors are simply turned off and thus
there is little voltage variation between nodes.
Hence, we do not consider any voltage measurements when the transmitter was
sending a 1-bit. Instead, we only consider those measured when it was sending a
0-bit, and refer to those as {\em dominant voltages}.

{\bf Variations in dominant voltages.}
From Fig.~\ref{fig:cantrans2}, when transceiver $i$ is transmitting a
0-bit, the current, $I_{(i)}$ flowing from its $V_{CC(i)}$ to its ground can be
derived as $I_{(i)} = \frac{{{V_{CC(i)}} - V_{G(i)} - 2{V_D}}}{{{R_{DSON,P(i)}}
		+ {R_{DSON,N(i)}} + {R_L}}}$,
where $V_{G(i)}$ denotes its ground voltage, and
$V_D$ the diodes' forward bias (assuming they are equivalent).
To simplify the analysis, we omit other factors such as leakage current or
variations in diodes.
We can thus derive the CANH and CANL dominant voltages,  $V_{CANH(i)}$ and
$V_{CANL(i)}$, from transceiver $i$ as:
\begin{equation}\label{eqn:vcan}
\setlength{\jot}{0pt} 
\begin{split}
&{V_{CANH(i)}} = {V_{CC(i)}} - {V_D} - {I_{(i)}}{R_{DSON,P(i)}},\\
&{V_{CANL(i)}} = {V_{G(i)}} + {V_D} + {I_{(i)}}{R_{DSON,N(i)}}.
\end{split}
\end{equation}
\noindent From Eq.~(\ref{eqn:vcan}), one can see that
\begin{itemize}
	\item[$\mathbb{V}2$.] \emph{Variations in $V_{CC}$, ground, and 
	$R_{DSON,P/N}$
		result in different ECUs with different CANH and CANL
		dominant voltages.}
\end{itemize}
%

\noindent For this reason, the ISO11898-2 specifies that a compliant
transceiver must accommodate dominant voltages
of CANH=2.75$\sim$4.5V and CANL=0.5$\sim$2.25V~\cite{iso2}.
Hence, we refer to any voltage values meeting this requirement as
{\em dominant voltages}.

{\bf Transient changes in on-state resistances.}
In Fig.~\ref{fig:canblock2}, when $V_{OUT+}$ of the driver increases,
$V_{OUT-}$ concurrently decreases as they are differential outputs. So, for
both transistors, the absolute differences between their gate and source
voltages simultaneously decrease. This results in both $R_{DSON,P}$ and
$R_{DSON,N}$ to increase, i.e., change in the {\em same}
direction~\cite{gatevolt, mosfet}.
Even when a change in the ECU temperature affects $R_{DSON,P}$ \& $R_{DSON,N}$,
they change in the same direction.
So, for a given $V_{CC}$ and ground voltage, the opposite signs of 
$I_{(i)}R_{DSON,P/N(i)}$ in
(\ref{eqn:vcan}) indicate that
\begin{itemize}
	\item[${\mathbb{V}3}$.] \emph{Transient changes in the ECU temperature and
		driver's input/output affect $R_{DSON,P/N}$, and thus make $V_{CANH}$ 
		and
		$V_{CANL}$ temporarily deviate in the ``opposite'' direction.}
\end{itemize}

%

Since regulated $V_{CC}$ and ground voltage remain constant, and are not
affected by transient changes in $R_{DSON,P/N}$,
\begin{itemize}
	\item[$\mathbb{V}4$.] \emph{Transient changes in $V_{CC}$ and ground are
		significantly smaller
		than those in $V_{CANH}$ and $V_{CANL}$, i.e., their values remain 
		relatively 
		constant.}
\end{itemize}

$\mathbb{V}1$--$\mathbb{V}4$ indicate that CANH and CANL dominant voltages
of each ECU are different from each other.  \name\ exploits this fact
in constructing different voltage profiles for (fingerprinting) ECUs.

\begin{figure}[t]
	\centering
	\includegraphics[width=\linewidth]{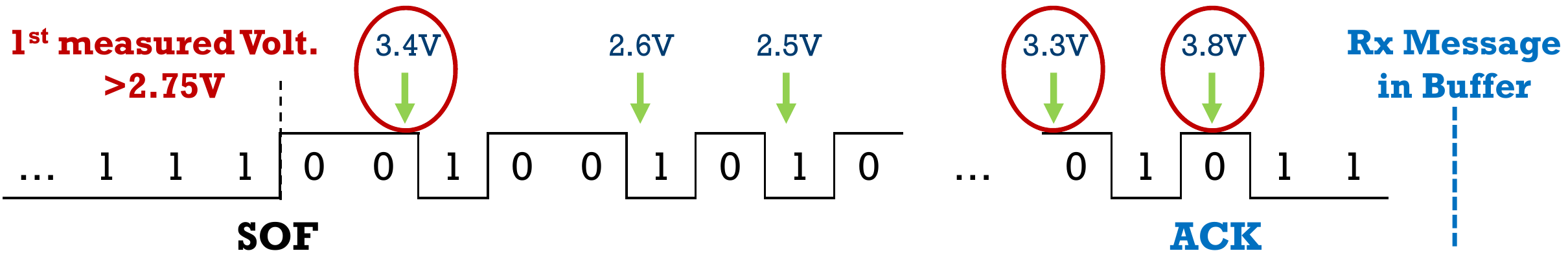}
	\caption{\name\ measuring CANH voltages.}
	\label{fig:ackthex}
\end{figure}


\subsection{Phase 1: ACK Threshold Learning}\label{sec:learnack}
\name\ is designed to run with a low voltage sampling
rate so that it can be easily installed as a low-cost software application,
which requires no changes in the CAN protocol; the high rate of
voltage sampling would only be required for the CAN protocol to receive
messages as it is designed to be.
Such a feature, however, renders \name\  incapable of determining at which
slot the voltage values were measured; all it knows is the value.
Thus, \name\ goes through a phase of learning the {\em ACK threshold}, which
determines whether or not the measured voltage was outputted by the message
transmitter.

{\bf Measuring dominant voltages.}
\name's measurement is triggered whenever a CANH voltage exceeds 2.75V
after a certain idle period.
This is because the first measured voltage exceeding 2.75V represents
the case of some transmitter transmitting a 0-bit on the bus~\cite{iso2}.
Since \name\ is only interested in dominant voltages, it discards any
measurements that are lower than 2.75V on CANH and higher than 2.25V on CANL.
The measurement continues until some message is shown to have been
received into \name's receive message buffer, i.e., an indication that
the transmitter has finished sending a message.
By reading the ID value of that received message, \name\ knows
which message ID the acquired dominant voltages correspond to.

{\bf Non-ACK voltages.}
\name\ continues collection of more dominant voltages for the acquired ID
(whenever the message is received) until
it learns its CANH and CANL ACK thresholds.
When collecting and exploiting voltage measurements, one needs to be cautious
of the fact ``During the ACK slot of a transmitted message, if received, all
other nodes but its transmitter output a 0-bit on the CAN bus''~\cite{canspec}.
Thus, even though \name\ samples at least a few dominant voltages while
receiving a certain message, {\em not all} represent the outputs from
the actual message transmitter.
Fig.~\ref{fig:ackthex} shows an example of \name's five voltage measurements of
\{3.4V, 2.6V, 2.5V, 3.3V, 3.8V\} from the CANH line during the reception of a
message, where 3.8V was measured during the ACK slot.
Of them, \name\ discards measurements \{2.6V, 2.5V\} as they do not meet the
criteria of dominant voltages.
If \name\ had considered the remaining 3 measurements as if they were output by
the message transmitter, it would have been incorrect since 3.8V was from all
ECUs but the message transmitter in the ACK slot.
Therefore, to accurately fingerprint the transmitter ECU, \name\ derives
the ACK threshold which distinguishes a non-ACK voltage measurement from an ACK
voltage measurement. We refer to {\em non-ACK voltages} as dominant voltages 
measured 
from slots other than the ACK slot, and {\em ACK voltages} as those measured 
from the ACK slot.
The threshold is derived by exploiting the following two facts of the ACK 
voltage.
\begin{itemize}
	\item[$\mathbb{K}1$.] {\em Low probability}:
	Since ACK is only 1 bit long, when measuring dominant voltages during a 
	message
	reception, most of them would be outputted from the message transmitter.
	\item[$\mathbb{K}2$.] {\em Different voltage level for ACK}:
	During an ACK slot, all ECUs but the transmitter acknowledge their message
	reception. Since those responders are connected in parallel and turned on
	concurrently, when receiving the ACK, the measured voltages are much higher 
	on CANH 
	and much lower on CANL than those when receiving non-ACK bits.
\end{itemize}
\name\ exploits these facts to collect $M$ dominant voltages from both CANH and 
CANL for
$N$ rounds for a given message ID.
So, based on $\mathbb{K}1$, the {\em most frequently} measured voltage value 
(of the $M$ values)
will most likely represent the non-ACK voltage.
During the $N$ rounds, we refer to the set of $N$ most frequently measured 
values as
the {\em most frequent set}, $S_{freq}$.
On the other hand, if we were to determine the {\em maximum} and the {\em 
minimum} of the
$M$ values from CANH and CANL, respectively, then they would represent ACK as 
well as
non-ACK voltages.
This is because even a single dominant voltage value collected (without 
awareness) from
the ACK slot would become the maximum/minimum of the $M$ values due to 
$\mathbb{K}2$.
Here, the set of $N$ maximum/minimum values measured from CANH/CANL is defined 
as the
{\em maximum/minimum set}, denoted as $S_{max/min}$.
For each message ID, \name\ exploits sets $S_{freq}$ and $S_{max/min}$
to derive the ACK threshold that differentiates a non-ACK voltage from an ACK 
voltage.

\begin{figure}[t]
	\centering
	\includegraphics[width=0.95\linewidth]{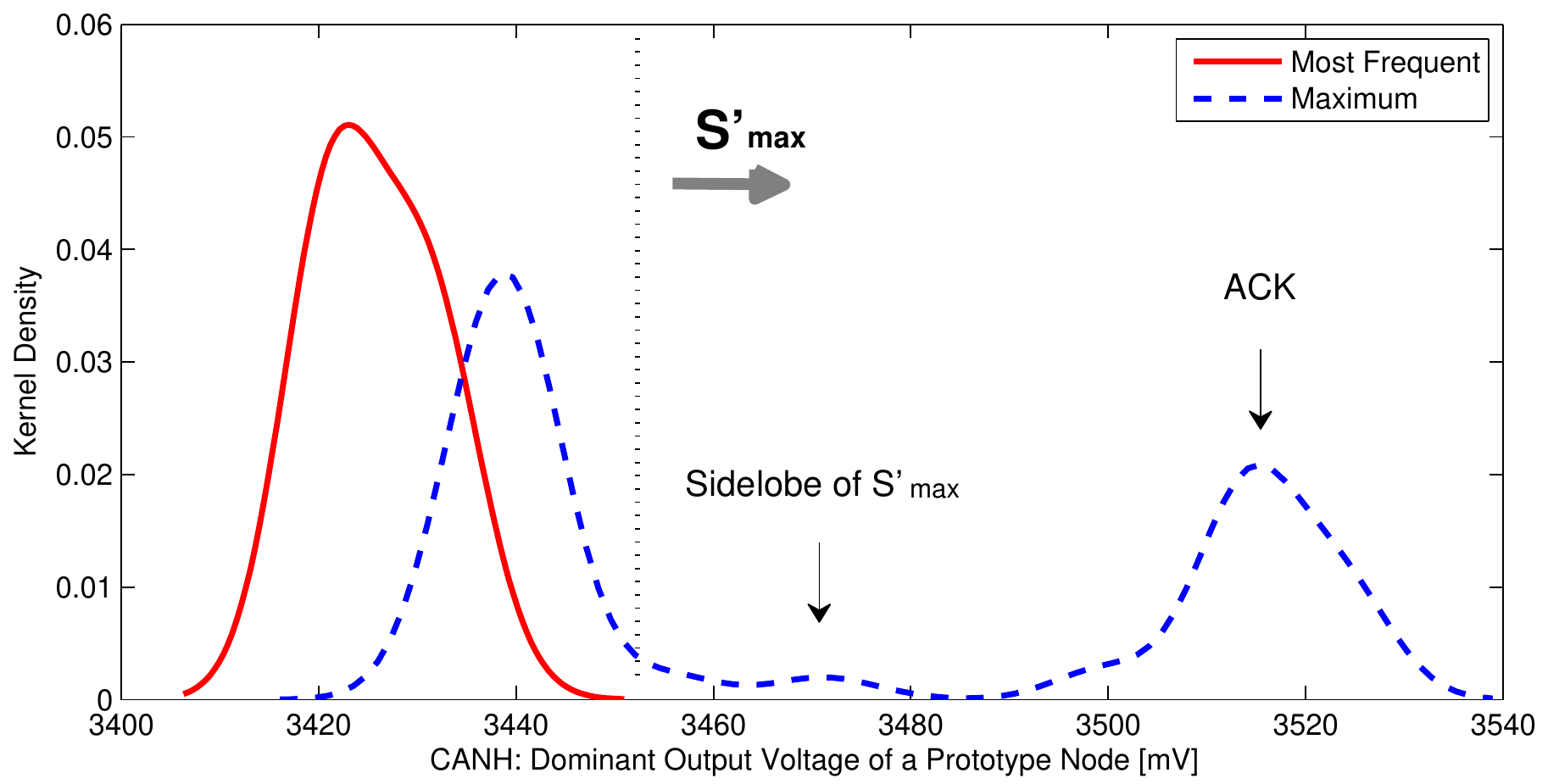}
	\caption{ACK threshold in a CAN bus prototype.}
	\label{fig:acknonack}
\end{figure}

{\bf Derivation of ACK threshold.}
Fig.~\ref{fig:acknonack} shows the kernel density plots of the most frequent and
the maximum sets of the measured dominant voltages from the CANH line.
The measurements were made while running \name\ on our CAN bus prototype, which
will be detailed in Section~\ref{sec:evaluation}.
One can see that only for the maximum set, there exists a side lobe, whereas
the most frequent set resembles a Gaussian distribution. Note that during
the $N$ rounds of $M$ measurements each, the most frequent and the maximum
values can be different. Thus, from the maximum set, \name\ first discards
values lower than $\max({S_{freq}}) + B{\sigma _{{S_{freq}}}}$, where
$\sigma_{{S_{freq}}}$ is the standard deviation of set $S_{freq}$, and $B$ a
design parameter determining how aggressive one wants to be in discarding ACK 
voltages.
Note that such a value also represents the rightmost end-point of the most 
frequent set's kernel density
(e.g., dotted vertical line in Fig.~\ref{fig:acknonack}).
Then, the usual side lobe of the maximum set ($S_{max}$) becomes the main lobe 
of
a refined maximum set, $S'_{max}$.
From $S'_{max}$, \name\ determines $\Gamma_1=median(S'_{max})-3MAD(S'_{max})$
and $\Gamma_2=\mu_{S'_{max}}-3\sigma_{S'_{max}}$, where $MAD(x)$ denotes the 
median
absolute deviation of $x$, and $\mu_{x}$ its mean. The CANH ACK threshold of 
the given message ID
(or its transmitter), $\Gamma_{ACK}^H$, is then derived to be $\max (\Gamma_1, 
\Gamma_2)$.
We take the maximum of the two to be conservative in discarding any non-ACK 
voltages.
Moreover, not only the lower 3$\sigma$ limit but also the lower 3-MAD limit is 
used since the refined
maximum set $S'_{max}$ may still contain its own (new) side lobe as shown in
Fig.~\ref{fig:acknonack}, i.e., an outlier for $S'_{max}$.
Using these processes, the ACK threshold of the example in 
Fig.~\ref{fig:acknonack} is
determined to be $\Gamma_{ACK}^H=3.499$V --- a point where
the two lobes in the maximum set are separated.
Depending on the transmitter ECU, the ACK threshold can be different as the set 
of responders
is different. Thus, the ACK learning is performed for all message IDs of 
interest.

When deriving the CANL ACK threshold, $\Gamma_{ACK}^L$, the minimum
(instead of the maximum) and the upper (instead of the lower) limits are used.
In Appendix~\ref{sec:acklearn_real}, we show that the proposed scheme can 
correctly 
determine the ACK thresholds even in real vehicles.

%
%

\subsection{Phase 2: Deriving a Voltage Instance}\label{sec:phase2}
Once ACK thresholds, $\Gamma_{ACK}^H$ and $\Gamma_{ACK}^L$, of the given
message ID are learned, from that point and on, \name\ continuously collects
dominant voltages, but discards those from CANH that are
lower than 2.75V or higher than $\Gamma_{ACK}^H$, and those from CANL that are
higher than 2.25V or lower than $\Gamma_{ACK}^L$.
This way, \name\ selects and processes only {\em non-ACK voltages}.
Whenever \name\ obtains $\kappa$ new measurements of CANH and CANL non-ACK
voltages, \name\ derives a new {\em voltage instance} which is defined as the
set of 6 tracking points, {\bf F}$_1$--{\bf F}$_6$.

{\bf F$_1$--F$_2$: Most frequent values.}
Similarly to Phase 1, 
\name\ determines the most frequently measured CANH and CANL voltages
(from $\kappa$ values), which are denoted as $F_1$ and $F_2$, respectively.
Since \name\ knows the ACK thresholds, the main differences from Phase 1 are
that only non-ACK voltages as well as $\kappa$ ($<M$) of them are used in
deriving the most frequent values. This way, \name\ keeps track of the {\em
	median} of the transmitter's dominant voltages.

\setlength{\textfloatsep}{11pt}
\begin{algorithm}[t]
	\small
	\caption{Dispersion Update}
	\label{alg:skewup}
	\begin{algorithmic}[1]
		\Function{UpdateDispersion}{$V, \Lambda, P^*$}
		\State \Return $\Lambda \gets \Lambda+ \alpha
		(P^*-\frac{\#(V<\Lambda)}{\#V})^3$ \Comment{Adjust tracking 		
			position}\label{penalty}	
		\EndFunction
		\If{\#measured CANH and CANL voltages both $\ge \kappa$}
		\State $V_H, V_L \gets \{$past $\kappa R$ CANH, CANL measurements$\}$
		\State $F_3 \gets \Call{UpdateDispersion}{V_H,F_3,0.75}$
		\State $F_4 \gets \Call{UpdateDispersion}{V_L,F_4,0.25}$
		\State $F_5 \gets \Call{UpdateDispersion}{V_H,F_5,0.9}$
		\State $F_6 \gets \Call{UpdateDispersion}{V_L,F_6,0.1}$
		\EndIf
	\end{algorithmic}
\end{algorithm}

{\bf F$_3$--F$_6$: Dispersions.}
\name\ also keeps track of the {\em dispersions} of CANH and CANL dominant 
voltages.
As the transmitter's voltage output behavior can change over time,
\name\ continuously updates 4 different {\em tracking points}, F$_3$--F$_6$, 
which
reflect (1) F$_3$: 75th,  (2) F$_5$: 90th percentile
of the transmitter's CANH outputs, (3) F$_4$: 25th, and (4) F$_6$:
10th percentile of CANL outputs.
By tracking the transmitter's voltage distribution, \name\ understands
its {\em momentary} voltage output behavior.
Thus, voltage instances represent those momentary behaviors.
Since even a single ACK voltage can significantly distort \name's
understanding of transmitters' behaviors, it is important to learn
the ACK threshold.
The reasons for \name's tracking of different percentiles of CANH and CANL are
that the low percentiles of CANH would contain voltages measured when the
transmitter switches from sending a 1-bit to sending a 0-bit, and vice versa.
The same applies for the high percentiles of CANL measurements. Although other
percentiles can be tracked as well, to minimize \name's overhead, we only track
F$_3$--$F_6$.

Algorithm \ref{alg:skewup} describes how the tracked dispersions are updated
whenever \name\ acquires $\kappa$ dominant voltages from each of CANH and CANL.
Using the past $\kappa R$ measurements, as in line \ref{penalty}, \name\
roughly estimates what percentile the current tracking point, $\Lambda$,
represents. In \name, we set $R=10$. Then, to correct and thus move the
tracking point $\Lambda$ to the desired position ---
where it represents the $P^*$ percentile --- an adjustment
is made as in line \ref{penalty}, where $\alpha$ is a design parameter
determining the sensitivity to changes.
With the adjustment function proportional to
$(P^*-\frac{\#(V<\Lambda)}{\#V})^3$, the tracking points move faster
if they are far away from their desired positions.
As a result, the four tracking points move if the transmitter's voltage 
distribution
(i.e., output behavior) shows changes, thus adapting to any changes on the CAN 
bus.
Instead of tracking, it is also possible to directly derive the
percentiles from the $\kappa R$ values. \name, however, does not follow this 
since it is too
sensitive to transient changes, especially when $\kappa R$ is small, i.e., 
insufficient
samples in deriving the percentiles. Thus, in order to make \name\ work under
various circumstances, we {\em track} them instead.

\subsection{Phase 3: Attacker Identification}\label{sec:voltprof}
A voltage instance (F$_1$--F$_6$) represents the momentary voltage output
behavior of the message transmitter. So, to log its usual behavior, \name\ 
exploits 
every newly derived voltage instance to construct/update the {\em voltage 
profile} 
of the message transmitter. Although the voltage instances are derived 
"per message ID", if messages originate from the same transmitter/ECU, their 
instances 
are near-equivalent, thus leading to construction of the same voltage profile.
We will later show through evaluations that there exists only one voltage
profile for a given transmitter/ECU, thus enabling its fingerprinting.
By exploiting a newly derived voltage instance, \name\ first updates the
{\em cumulative voltage deviations} (CVDs) of features F$_1$--F$_6$.
We define a CVD to represent how much the transmitter's dominant
voltages deviated cumulatively from their ideal values. Thus, for feature
F$_x$, the CVD at step $n$, $CVD_x[n]$, is updated as:
\begin{equation}\label{eqn:cvd}
{CVD}_x[n] = CV{D_x}[n-1] + {\Delta[n]}\left( {1 - {\nu_x[n]}/{\nu_x^*}} 
\right),
\end{equation}
\noindent where $\Delta [n]$ is the elapsed time since step $n-1$, $\nu_x[n]$ 
the value
of feature $F_x$ at step $n$, and $\nu^*_x$ the desired value of $\nu_x$.
Ideally, the most frequently measured as well as any percentiles of the CANH and
CANL dominant voltages should be equal to 3.5V and 1.5V, respectively, i.e.,  no
variations in their output voltages.
Therefore, for features \{F$_1$, F$_3$, F$_5$\}, which represent CANH values, 
we set $\nu^*_{\{1,3,5\}}=3.5V$ and similarly we set $\nu^*_{\{2,4,6\}}=1.5V$.

{\bf Suppressing transient changes.}
As ECUs have different $V_{CC}$, ground, and $R_{DSON}$ values, they output
different CANH and CANL dominant voltages.
Their momentary voltage instances would, therefore, be different, and hence
the trends in their CVD changes would also be different from each other.
So, for every obtained CVD of features F$_1$--F$_6$, \name\ derives
$\Psi [n]= \sum_{x = 1}^6 {CV{D_x}[n]}$.
The reason for \name's summing of all the CVDs is to exploit $\mathbb{V}3$.
Recall from Section~\ref{sec:facts} that $\mathbb{V}3$ gives us transient
deviations in CANH and CANL output voltages are opposite in direction.
So, via CVD summation, \name\  {\em suppresses} any transient
deviations that have occurred (due to changes in driver, temperature, etc.)
when constructing and/or updating the voltage profiles.
Note that since CAN is a differential bus, F$_2$, F$_4$, F$_6$
suppress F$_1$, F$_3$, F$_5$, respectively.

{\bf Voltage profile.}
Suppression of transient changes yields a value, $\Psi$, that (mostly)
represents the {\em consistent} factors in the voltage instances:
$V_{CC}$, ground voltages, and the usual voltage drops across the transistors.
As stated in $\mathbb{V}4$, since these values are rather constant, the 
accumulated
sum of $\Psi$, $\Psi_{accum}[n]=\sum_{k=1}^n \Psi[k]$ becomes linear in time.
Moreover, from $\mathbb{V}1$--$\mathbb{V}2$, as $\Psi$ values are distinct for
different ECUs,
the {\em trends} in how $\Psi_{accum}$ changes also become different, i.e., the
slopes in a $\Psi_{accum}$--time graph are different.
Therefore, \name\ formulates a linear parameter identification problem as
$\Psi_{accum}[n] = \Upsilon [n] t[n] + e[n]$, where
at step $n$, $\Upsilon[n]$ is the regression parameter, $t[n]$ the elapsed
time, and $e[n]$ the identification error.
As the regression parameter $\Upsilon$ represents the slope of the linear model
and varies with the transmitter, we define this as the {\em voltage profile}.
This way of formulating the problem and constructing the profiles facilitates 
\name's online
update of fingerprints, which is key to \name's adaptability.
To determine the voltage profile $\Upsilon$, i.e., fingerprint ECUs, we use an
adaptive signal processing technique, the Recursive Least Squares 
(RLS)~\cite{rls}, which 
is an online approach in learning the regression parameter.
Note, however, that the choice of algorithm does not affect \name's performance.
In RLS, we use kiloseconds $($=$10^3$ secs$)$ as the unit for $t$.
Due to space limitation, we omit details of RLS, and refer the readers to 
\cite{rls} for its details.
We will later show, via experimental evaluations, that the thus-derived
profile $\Upsilon$ is constant over time and also distinct for
different ECUs, thus allowing \name\ to correctly fingerprint them.

{\bf Identifying the attacker.}
When an adversary mounts an attack, the underlying IDS can determine whether
the message is malicious or not, so \name\ can filter out the voltage outputs
obtained only from the (detected) attack messages and build a voltage profile
from only those. We refer to such a voltage profile as an {\em intrusion
	voltage profile}. \name\ then looks up the voltage profiles it had built
until the detection of the attack and searches for the one that is similar
to the intrusion voltage profile.\footnote{The initial set of ``ground
	truth'' voltage profiles can be verified via timing-based fingerprinting
	schemes~\cite{dinatale,cids}.}
This way, \name\ identifies the attacker ECU.

The performance of \name\ will, of course, depend on how well the 
IDS detects the intrusion; this dependency needs to be investigated when 
an IDS and \name\ are integrated as a whole system. 
Note, however, that the mostly periodic nature of 
in-vehicle messages makes correct detection of intrusions not as 
difficult as pinpointing the attacker ECU. Researchers and 
car-makers are now well aware of how to detect intrusions, 
but not how to accurately identify the attacker ECU.

The only case where the identified ECU would have an
unknown/unlearned profile is when it was physically attached to the vehicle by
an adversary. However, since this requires physical access and its
identification has been addressed elsewhere \cite{busoff2,voltarxiv}, we do not
discuss its detection any further in this paper.

\subsection{Phase 4: Verification}\label{sec:verf}
By the birthday paradox, two different ECUs may naturally have near-equivalent
voltage profiles, i.e., voltage profile collision, thus confusing \name\
in identifying the attacker ECU.
Note, however, that \name\ has at least narrowed its search scope significantly.
An adversary may also attempt to mimic some other
ECU's voltage output behavior, i.e., targeted impersonation.
In such a case where further verification besides the voltage profiles is
required, in Phase 4 of \name, machine classifiers are run with the (momentary)
{\em voltage instances} as their inputs, i.e., F$_1$--F$_6$ as their features.
This way, an analysis of attacks from a different vantage point --- not only
its trend (Phase 3) but also its momentary behavior --- is performed, thus
resolving ambiguities in attacker identification.
We, however, stress that while the adaptability achieved from Phase 3 is an
essential attribute for an accurate attacker identification, Phase 4 cannot
totally replace it, i.e., only complements Phase 3.
We will later show through evaluations that by using
voltage instances as machine classifiers' input, \name\ can resolve issues such
as voltage profile collision and an adversary's targeted impersonation.

\subsection{Voltage Profile Adjustment}\label{sec:adjust}
For attacker identification, it is important to not only have the correct
fingerprint of an ECU but also that fingerprint to be still valid when
examining a voltage measurement obtained during the (detected) attack.
If it was updated much earlier than when the attack was
detected, any changes occurred between those two time instants would
not be reflected in the latest model, thus leading to false identifications.
We refer to this as a {\em model-exam discrepancy}.
Since \name\ continuously updates the voltage profiles in real time,
such a model-exam discrepancy is minimized/nullified.
Since attacker identification is performed {\em upon} detection of an
intrusion, as long as \name\ keeps the fingerprints up-to-date until an
intrusion is detected (by an IDS), \name\ can locate the source of the attack.
Even when there are abrupt changes in the temperature of an ECU, \name\
suppresses those transient changes, and adapts its model accordingly for an
accurate attacker identification.

One corner case in which the performance of \name\ might suffer from
the model-exam discrepancy would be when the vehicle has not been turned on
for a long time. During that period, various features (e.g., power supply
level, ambient temperature) which affect the output voltages might have
changed. In such a case, since the old voltage profiles may not correctly
reflect the current status, \name\ may have to reconstruct (instead of update)
them. In fact, a timing-voltage-aware adversary may attempt to exploit such a
fact and attack the CAN bus as soon as the vehicle is turned on,
making \name\ incapable of handling the attacks. However, even in such a case,
as ECUs use the same power source, i.e., battery, and thus all voltage profiles
change in the same direction and with the same magnitude, \name\ re-adjusts and
reuses the old ones as a starting point for voltage profile {\em update} rather
than reconstructing it from scratch when the vehicle is turned on.
Specifically, \name\ first determines how much of {\em common} changes
occurred in ECUs' $V_{CC}$ by deriving the differences between the previous and
current mean values of $($F$_3$+F$_4$+F$_5$+F$_6)/2$ --- an estimated
value of $V_{CC}$ based on Eq.~(\ref{eqn:vcan}).
\name\ then adds the thus-derived differences to $\nu^*$ (in
Eq.~(\ref{eqn:cvd})) based on the fact that if common changes in $V_{CC}$
incur, CANH and CANL output values increase simultaneously~\cite{ti1,ti2}.
This way, \name\ correctly adjusts/updates its voltage profile(s) and thus
identifies such a type of timing-voltage-aware adversary; we will later
evaluate this via real vehicle experiments.
Note, however, that if voltage-based fingerprinting was done solely via batch
learning (as in \cite{sourceiden,voltarxiv}), it cannot make such an
adjustment, suffer from high model-exam discrepancy, thus allowing a
timing-voltage-aware adversary to evade it.

\subsection{Security of \name}\label{sec:secfeat}
Once an intrusion is detected, via voltage measurements, \name\ can identify
the attacker ECU.

A {\em naive} adversary would be capable of controlling the vehicle via
continuous message injections. However, since he has no knowledge of
how ECUs might be fingerprinted, he would inject them imprudently.
In such a case, he cannot evade \name.

A {\em timing-aware} adversary who knows that ECUs can be fingerprinted via
timing analysis, will attempt to exploit this knowledge in not only
controlling the vehicle but also evading the fingerprinting device.
For example, the adversary may know that CIDS~\cite{cids} can identify the 
attacker ECU
only if the attack messages were injected periodically.
Hence, he may perform an arbitrary impersonation by injecting messages
{\em aperiodically}, thus fooling CIDS. Note, however, that CIDS
would still detect the presence of the attack. In addition, based on his
knowledge that CIDS's fingerprints are basically clock skews, he may
attempt to imitate the targeted ECU's clock behavior, i.e., targeted
impersonation. However, with \name\ also installed in the vehicle, since it
identifies the attacker ECU via voltage measurements, i.e.,
irrespective of message timings, a timing-aware adversary can evade CIDS, but
not \name.

A {\em timing-voltage-aware} adversary may also try to evade \name\ using
his knowledge of how voltage-based fingerprinting devices run.
In order to achieve this, when or before the adversary injects the attack 
messages, he may
attempt to change the voltage output levels by changing the supply voltage
(e.g., run processes which drain battery) or by heating up or cooling down the
ECUs so that the transistors' internal resistance values change.
He could even attempt to start attacking the CAN bus only when the vehicle is
turned on after staying off for a long time as discussed in
Section~\ref{sec:adjust}. However, since \name\ performs an online
update of voltage-based fingerprints and also adjusts them if
necessary, thus minimizing/nullifying model-exam discrepancy, it would be
difficult for the timing-voltage-aware adversary to evade \name.
Moreover, since \name\ analyzes voltage outputs from two different
perspectives --- momentary behavior (Phase 4) and its trend (Phase 3) ---
a timing-voltage-aware adversary incapable of precisely controlling the
instantaneous voltage outputs cannot evade \name.

\section{Evaluation}\label{sec:evaluation}
We now evaluate the practicability and efficiency of \name\ in
achieving an effective and accurate attacker identification on a 
CAN bus prototype and two {\em real} vehicles.
When running \name\ for both evaluation settings, in Phase 1, $M=30$ dominant 
voltages were obtained for each message ID for $N=50$ rounds. 
From Phase 2, voltage instances were outputted whenever 
$\kappa=15$ non-ACK voltages from both CANH and CANL were acquired.

\begin{figure*}
	\centering
	\begin{subfigure}[t]{0.45\linewidth}
		\centering
		\includegraphics[height=3.2cm]{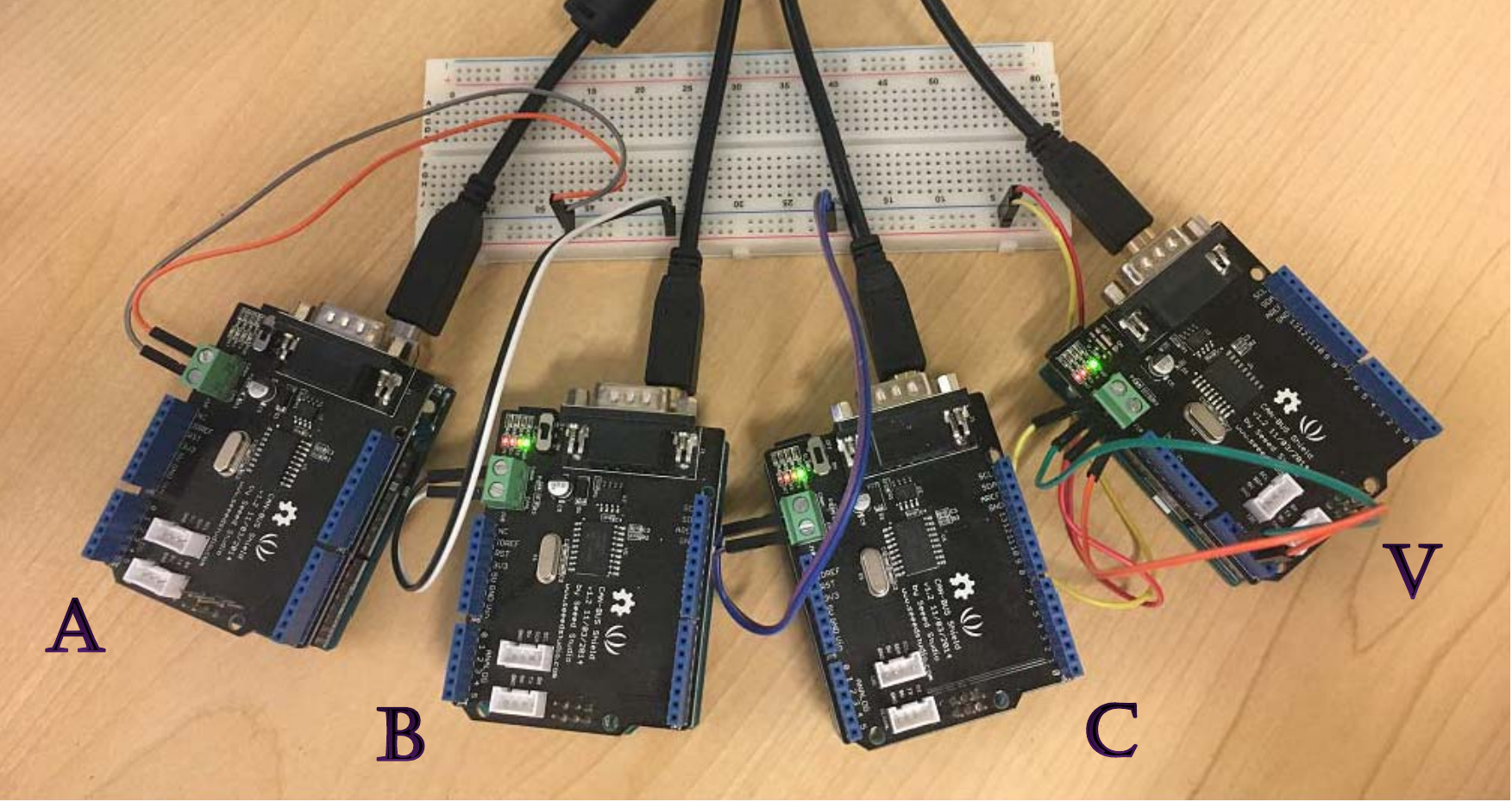}
		\caption{CAN bus prototype.}
		\label{fig:testbed}
	\end{subfigure}
	\hspace{1cm}
	\begin{subfigure}[t]{0.45\linewidth}
		\centering
		\includegraphics[height=3.5cm]{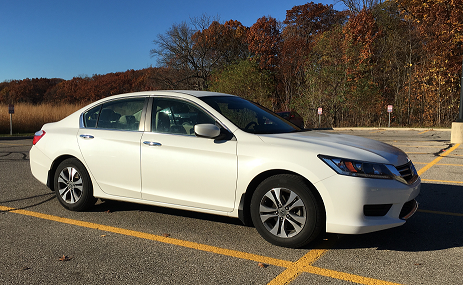}
		\caption{2013 Honda Accord.}
		\label{fig:vehicle}
	\end{subfigure}	
	\vspace{0.1cm}
	\begin{subfigure}[t]{0.45\linewidth}
		\centering
		\includegraphics[height=3.5cm]{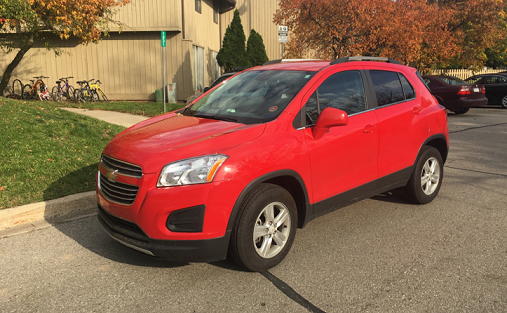}
		\caption{2015 Chevrolet Trax.}
		\label{fig:vehicle_chevy}
	\end{subfigure}	
	\hspace{1cm}
	\begin{subfigure}[t]{0.45\linewidth}
		\centering
		\includegraphics[height=3.5cm]{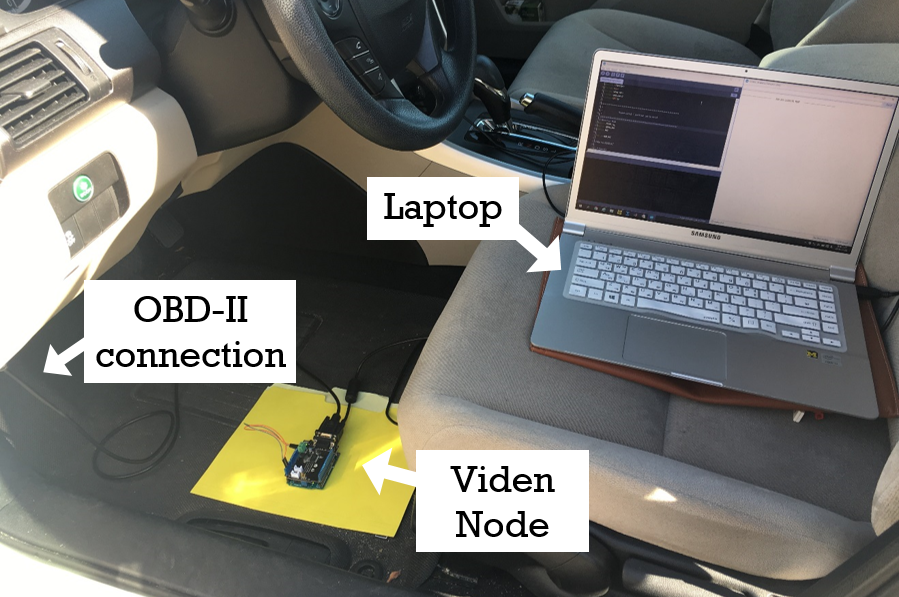}
		\caption{Connection to the vehicle.}
		\label{fig:obdconn2}
	\end{subfigure}	
	\caption{Experiments were conducted on a CAN bus prototype and on two real
		vehicles.}
\end{figure*}

\subsection{Evaluation Setups}
{\bf CAN bus prototype.} 
As shown in Fig.~\ref{fig:testbed}, we configured a CAN prototype in which all 
four 
nodes were connected to each other.
Each node consists of an Arduino UNO board and a SeeedStudio CAN shield.
The CAN bus shield consists of a Microchip MCP2515 CAN controller 
and a MCP2551 CAN transceiver to provide CAN bus communication capabilities.
Only two nodes were configured to have a 120$\Omega$ terminal 
resistor so as to match $R_L=60\Omega$.

The three prototype nodes $\mathbb{A}$, $\mathbb{B}$, and $\mathbb{C}$ were 
programmed to inject messages 0x01, 0x07, and 0x15 at random message intervals 
within $[$20ms, 200ms$]$. 
The fourth node $\mathbb{V}$ was programmed to run {\name} and
construct voltage profiles for messages 0x01, 0x07, and 0x15 (i.e., 
transmitters $\mathbb{A}$, $\mathbb{B}$, and $\mathbb{C}$), respectively.
The reason for injecting the messages {\em aperiodically} is to show that even 
in such cases, \name\ is capable of fingerprinting the transmitters. 
For node $\mathbb{V}$ that runs \name, its CANH and CANL lines were not 
only connected to the bus but also to the
microcontroller's Analog-to-Digital Converter (ADC), which had 10-bit 
resolution and was configured to sample
voltages at its maximum rate of 50 KSamples/sec.
This way, $\mathbb{V}$ acquired measurements of dominant 
voltages on the bus when nodes $\mathbb{A}$--$\mathbb{C}$ were 
sending their messages.
The CAN bus prototype was set up to operate at 500Kbps, which is typical 
for in-vehicle high-speed CAN buses. 
In such settings, \name\ required only 2--3 messages to output a voltage 
instance and update the profiles. 

\begin{figure}
	\centering
	\begin{subfigure}[t]{0.8\linewidth}
		\centering
		\includegraphics[width=8cm]{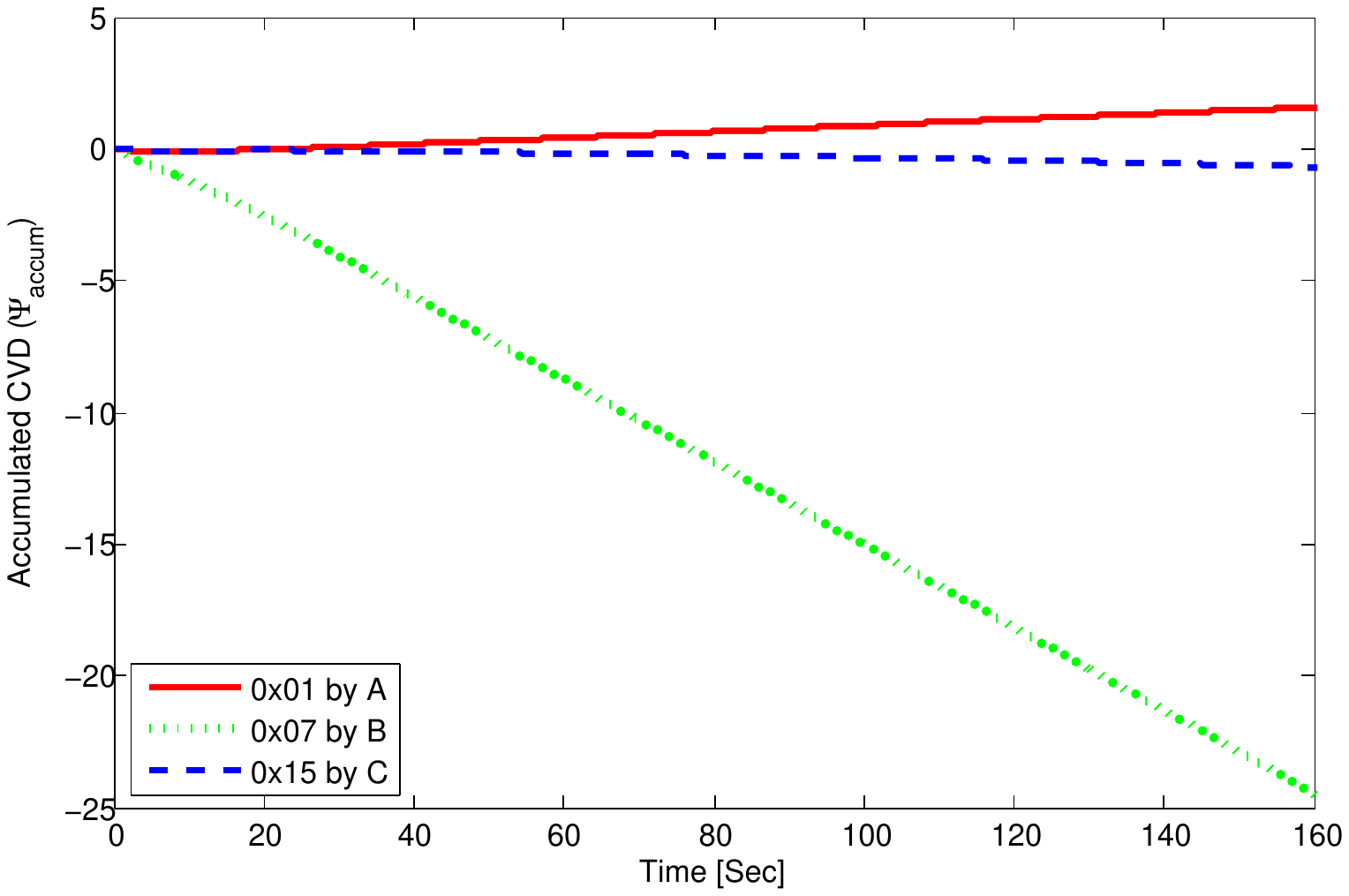}
		\caption{CAN bus prototype.}
		\label{fig:finger_proto}
	\end{subfigure}
	\begin{subfigure}[t]{0.8\linewidth}
		\centering
		\includegraphics[width=8cm]{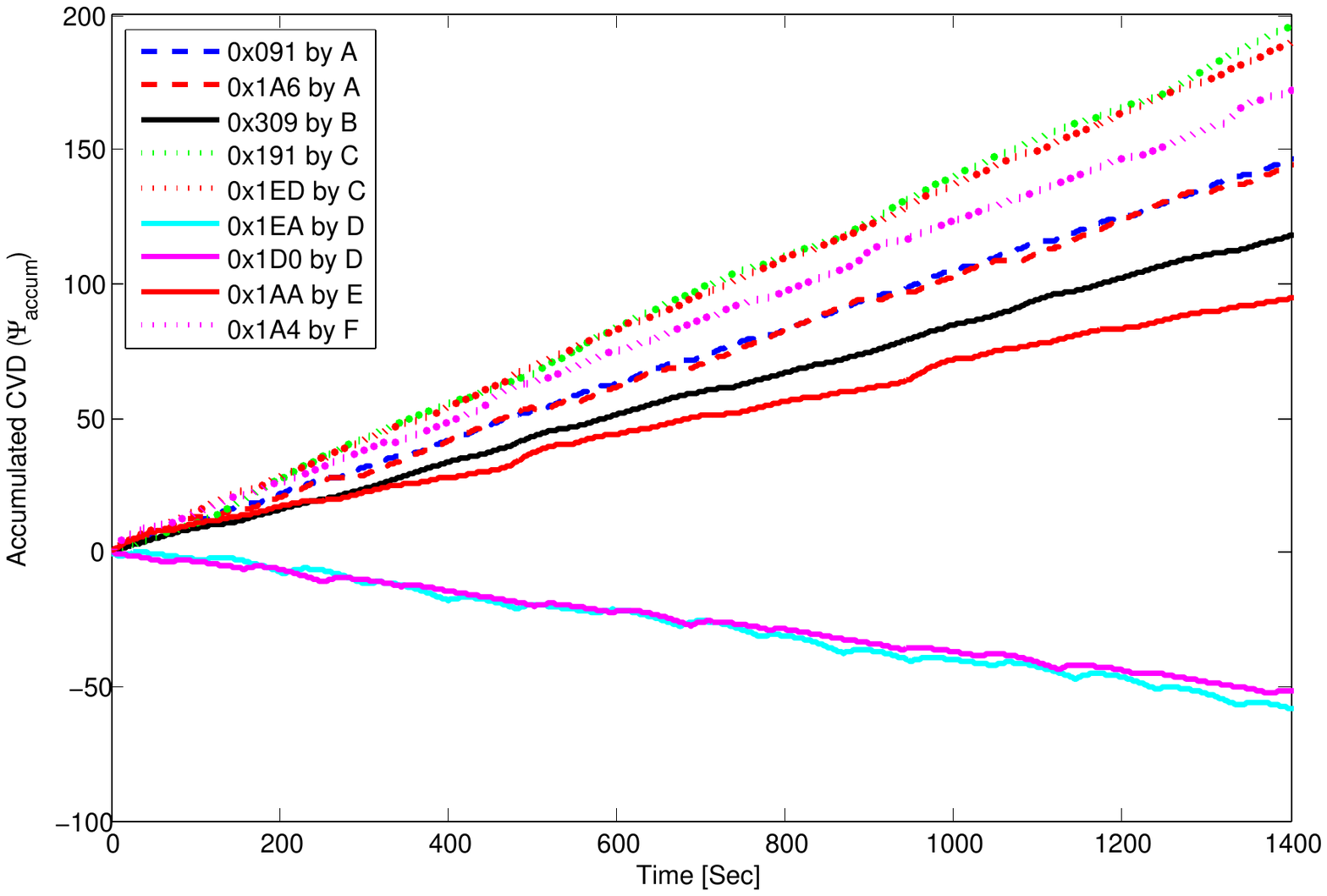}
		\caption{2013 Honda Accord.}
		\label{fig:finger_real}
	\end{subfigure}	
	\begin{subfigure}[t]{0.8\linewidth}
		\centering
		\includegraphics[width=8cm]{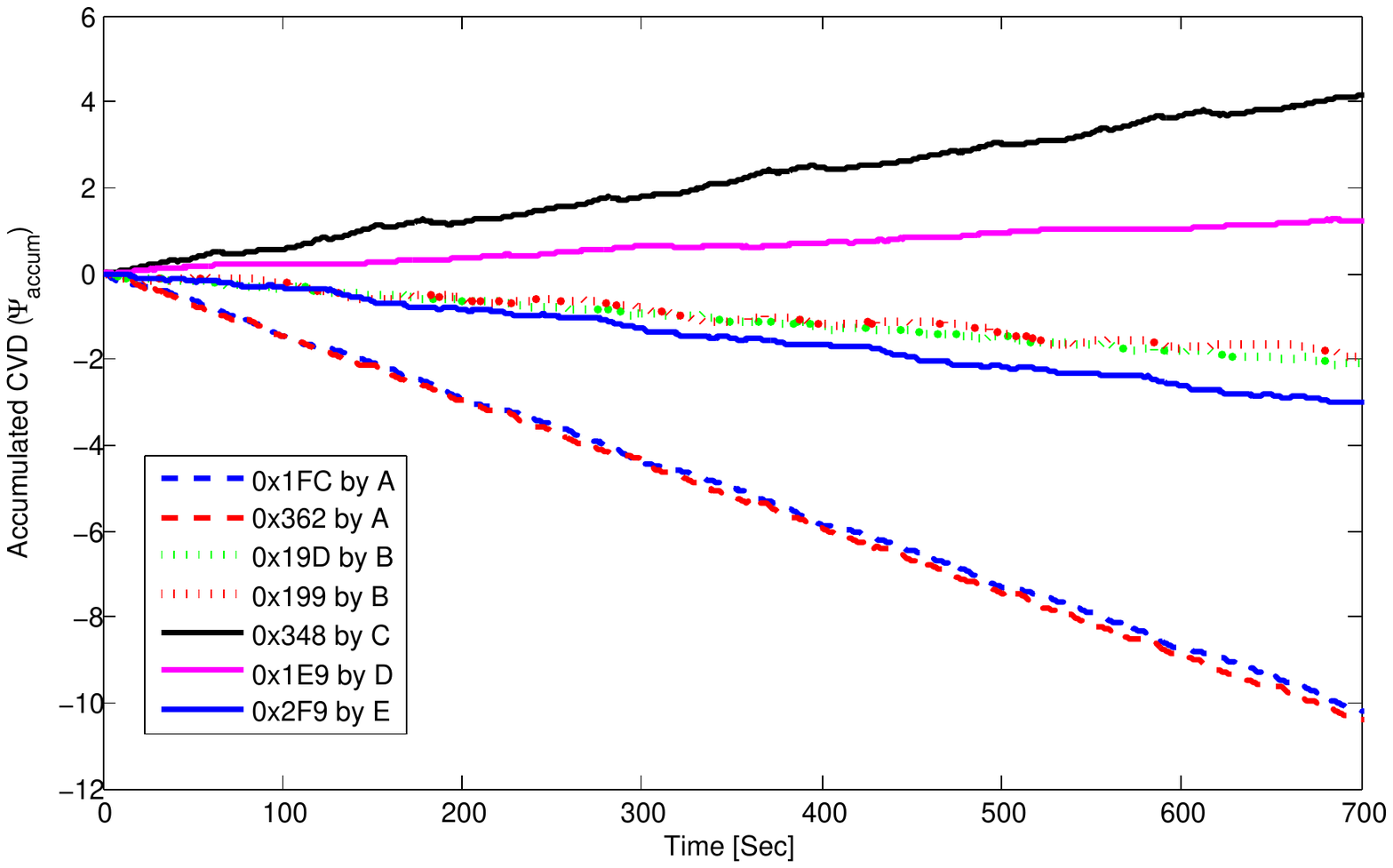}
		\caption{2015 Chevrolet Trax.}
		\label{fig:finger_real_chevy}
	\end{subfigure}	
	\caption{Voltage profiles obtained from the CAN bus prototype and the two
		real vehicles.}\label{fig:finger}
\end{figure}

{\bf Real vehicles.}
Two cars, 2013 Honda Accord (Fig.~\ref{fig:vehicle}) and a 2015 Chevrolet Trax 
(Fig.~\ref{fig:vehicle_chevy}),
were also used for our experimental evaluation of \name. Through the OBD-II 
port, 
the \name\ node ($\mathbb{V}$) was connected to the in-vehicle CAN 
bus, both running at 500Kbps.
From a laptop and through the \name\ node, as shown in Fig.~\ref{fig:obdconn2}, 
we were able 
to read messages from the 2013 Honda Accord's and the 2015 Chevrolet Trax's CAN 
buses.
While \name\ was receiving messages from the two vehicles, it sampled 
their CANH and CANL voltages and then derived their ECUs' voltage instances and 
profiles.

\subsection{Voltage Profiles as Fingerprints}
We first evaluate the accuracy and validity of using voltage profiles
to fingerprint the transmitter ECUs.

{\bf CAN bus prototype.}
Fig.~\ref{fig:finger_proto} shows the voltage profiles of all the three 
messages sent on 
the prototype bus. Although the three CAN prototypes nodes were built with 
the same hardware, the corresponding message IDs 
showed different trends in how their $\Psi_{accum}$ changed over time, since 
the three ECUs differ in their supply and transistor characteristics. 
Based on the RLS implemented in \name, we were able to find that nodes 
$\mathbb{A}$, $\mathbb{B}$, and $\mathbb{C}$ had different voltage profiles 
($\Upsilon$) 
being equal to 10.1, -154.3, and -4.9, respectively.
In other words, voltage profiles of 0x01, 0x07, and 0x15 were shown to 
be different from each other as they were sent by different ECUs, thus verifying
the feasibility and accuracy of \name.

{\bf Real vehicles.}
In the CAN prototype, we knew which ECU is sending which message(s),
but it is difficult to know this in a real vehicle.
In order to obtain the ground truth on the message source(s),
we exploit the schemes in \cite{dinatale, cids}, which analyzed timing patterns 
in CAN for fingerprinting the ECUs. Note, however, that these are
used only for obtaining the ground truth, since those cannot identify the 
attacker ECU if messages are injected at random times.

Through the connected \name\ node, we not only logged the CAN traffic of the 
2013 Honda Accord but also measured the dominant voltages from its CAN bus.
The measurements were made on a stationary vehicle, but while continuously 
changing
their operations (e.g., pressing brake pedal, turning the steering wheel) to 
generate some 
transient changes.
In Appendix~\ref{sec:whiledrive}, we show that outputs in \name\ is {\em not}
affected by whether the car is being driven or stationary.
By logging the CAN traffic and exploiting the schemes in \cite{dinatale, cids},
we were able to verify that messages 
\{0x091, 0x1A6\} were sent from some ECU $\mathbb{A}$, 
\{0x309\} from $\mathbb{B}$, \{0x191, 0x1ED\} from $\mathbb{C}$, 
\{0x1EA, 0x1D0\} from $\mathbb{D}$, \{0x1AA\} from $\mathbb{E}$, and
\{0x1A4\} from $\mathbb{F}$.
Fig.~\ref{fig:finger_real} shows the messages' voltage profiles.
The profiles ($\Upsilon$) derived by \name\ are shown to be equivalent 
{\em only} for those messages sent from the same ECU;
ECU $\mathbb{A}$ sending \{0x091, 0x1A6\} had $\Upsilon_{\mathbb{A}}=102.6$,
$\mathbb{B}$ sending \{0x309\} had $\Upsilon_{\mathbb{B}}=85.0$,
$\mathbb{C}$ sending \{0x191, 0x1ED\} had $\Upsilon_{\mathbb{C}}=137.0$, 
$\mathbb{D}$ sending \{0x1EA, 0x1D0\} had $\Upsilon_{\mathbb{D}}=-39.2$,
$\mathbb{E}$ sending \{0x1AA\} had $\Upsilon_{\mathbb{E}}=67.5$, while
$\mathbb{F}$ sending \{0x1A4\} had $\Upsilon_{\mathbb{F}}=120.8$.
This result again shows that voltage profiles for {\em different} ECUs are 
different and can thus be used as their fingerprints.

To further verify that \name's capability of fingerprinting is not restricted 
to a specific vehicle model, \name\ was also run on a 2015 Chevrolet Trax. 
Again, by exploiting the schemes in \cite{dinatale, cids}, we obtained the 
ground truths of messages 
\{0x1FC, 0x362\} sent from some ECU $\mathbb{A}$, 
\{0x19D, 0x199\} from $\mathbb{B}$, 
\{0x348\} from $\mathbb{C}$, \{0x1E9\} from $\mathbb{D}$, 
and \{0x2F9\} from $\mathbb{E}$.
Fig.~\ref{fig:finger_real_chevy} shows the result of \name\ determining that 
\{0x1FC, 0x362\} have a voltage profile of $\Upsilon_{\mathbb{A}}=-14.7$, 
\{0x19D, 0x199\} have $\Upsilon_{\mathbb{B}}=-2.8$, 
\{0x348\} has $\Upsilon_{\mathbb{C}}=5.9$, 
\{0x1E9\} has $\Upsilon_{\mathbb{D}}=1.8$, 
and \{0x2F9\} has $\Upsilon_{\mathbb{E}}=-4.4$. 
Thus, using voltage measurements, \name\ correctly fingerprinted their 
transmitters.
This again confirms the diversity of voltage profiles (of different ECUs), thus 
facilitating \name's fingerprinting of in-vehicle ECUs. Moreover, these results 
show that \name's fingerprinting is not limited to a specific vehicle model, and
can thus be applied to other vehicle models.

\begin{figure*}[!t]
	\centering
	\begin{subfigure}[t]{0.48\linewidth}
		\centering
		\includegraphics[width=0.95\linewidth]{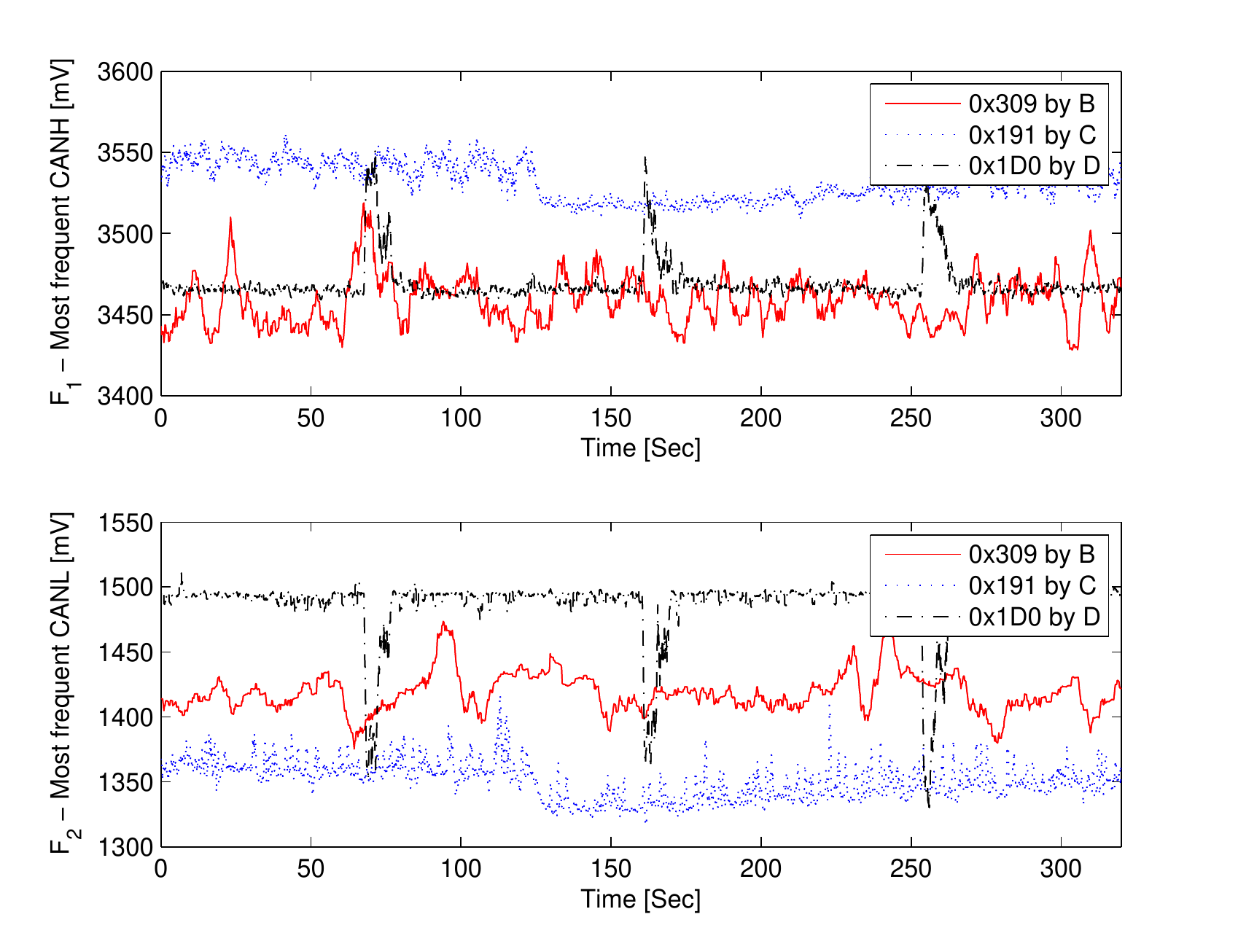}
		\caption{2013 Honda Accord.}
		\label{fig:v1v3}
	\end{subfigure}	
	~
	\begin{subfigure}[t]{0.48\linewidth}
		\centering
		\includegraphics[width=\linewidth]{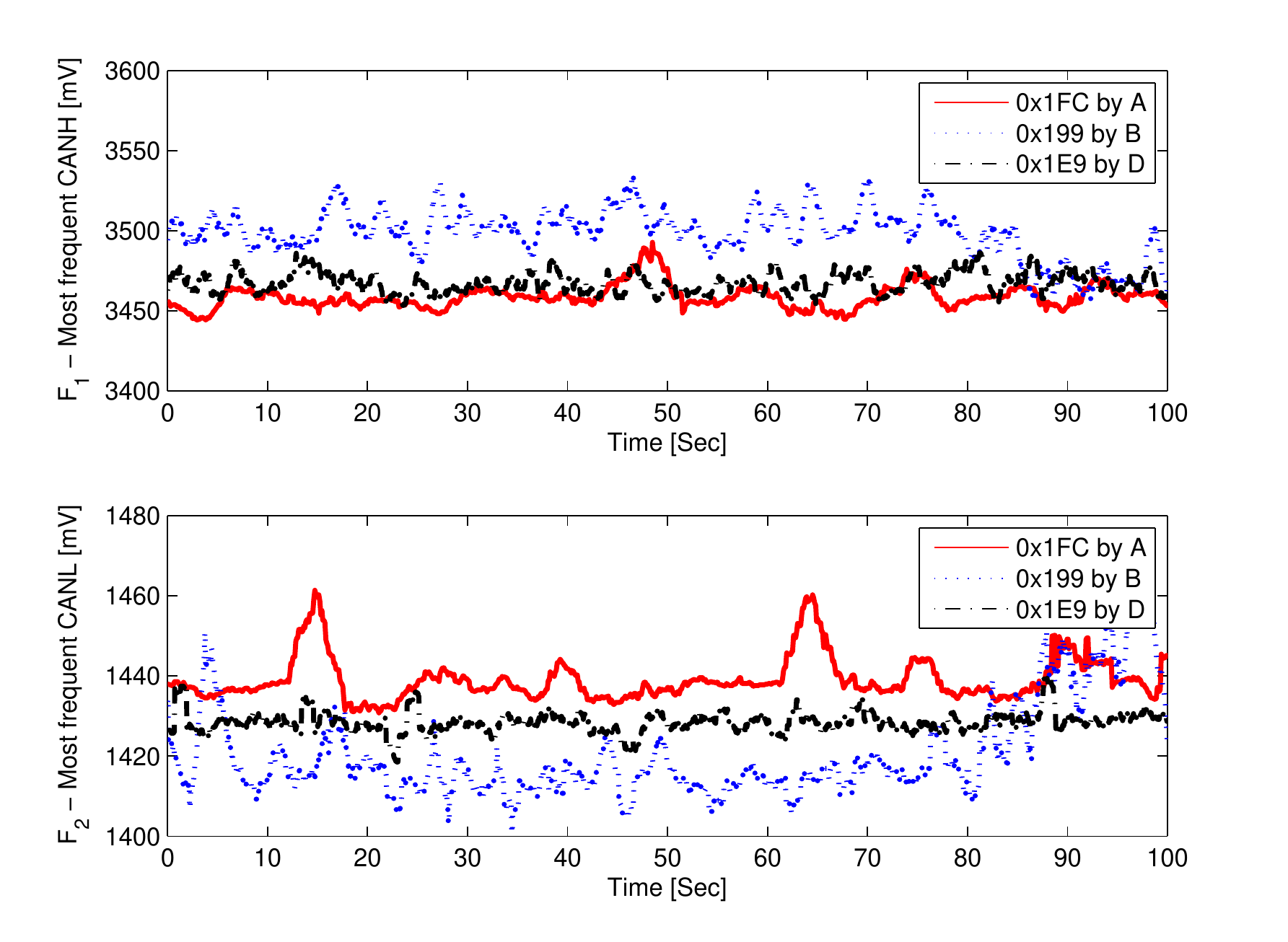}
		\caption{2015 Chevrolet Trax.}
		\label{fig:v1v3_chevy}
	\end{subfigure}	
	~
	\caption{Features F$_1$ and F$_2$ of \name\ in the two real vehicles.}
\end{figure*}

\begin{figure}[t]\centering
	\includegraphics[width=0.8\linewidth]{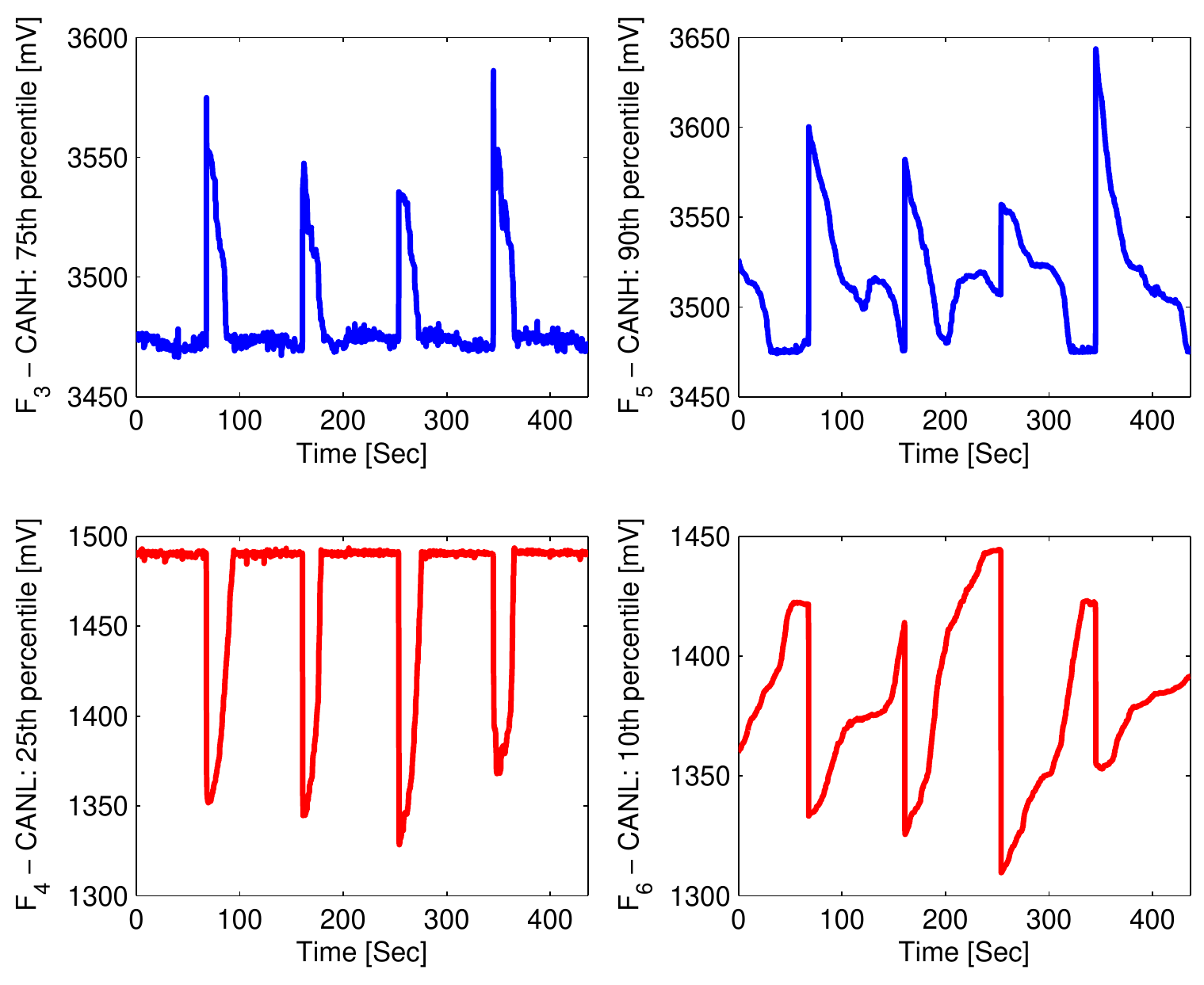}
	\caption{Changes of message 0x1D0 in the Honda Accord.}
	\label{fig:v4}
\end{figure}

\subsection{Voltage Outputs in Real Vehicles}
We provided 4 characteristics, $\mathbb{V}1$--$\mathbb{V}4$, which
were imperative for \name\ in fingerprinting ECUs. 
We evaluate whether $\mathbb{V}1$--$\mathbb{V}3$ actually hold in real vehicles.
Note that Fig.~\ref{fig:finger} verifies $\mathbb{V}4$, corroborating 
that the voltage profiles of ECUs were constant over time, i.e., linear.

{\bf Different outputs.}
According to $\mathbb{V}1$--$\mathbb{V}2$,  ECUs output different 
dominant voltages.
Fig.~\ref{fig:v1v3} plots features F$_1$--F$_2$ (i.e., the most frequently 
measured CANH and CANL values) 
outputted by \name\ for messages 0x309 (sent by $\mathbb{B}$), 0x191 (sent by 
$\mathbb{C}$), and 
0x1D0 (sent by $\mathbb{D}$) in the Honda Accord. 
Although the transceivers of all those messages are to output the agreed-on 
CANH=3.5V and CANL=1.5V when sending a 0-bit, they outputted values deviating 
from them. More importantly, their output levels were clearly discriminable.
Even though ECU $\mathbb{B}$, which sent 0x309, was shown to output similar 
CANH dominant voltages to ECU $\mathbb{D}$, it outputted totally different 
voltages on CANL. 
Similarly, Fig.~\ref{fig:v1v3_chevy} plots F$_1$--F$_2$ values of 0x1FC (sent 
from $\mathbb{A}$), 0x199 (sent from $\mathbb{B}$), and 0x1E9 (sent from 
$\mathbb{D}$) outputted by \name\ in the 2015 Chevrolet Trax.
Again, we can see that the transmitters of those messages did not output the 
desired levels, but outputted discernible levels. These results confirm that 
$\mathbb{V}1$--$\mathbb{V}2$ hold even in real vehicles, thus facilitating 
\name's fingerprinting.

{\bf Transient changes.}
$\mathbb{V}3$ states that transient changes in CANH and CANL voltages are 
opposite in direction.
Fig.~\ref{fig:v4} shows  the 4 tracked percentiles, F$_3$--F$_6$, of message 
0x1D0 in the 2013 Honda Accord. F$_3$--F$_6$ values are shown to temporarily 
deviate from and later return to their usual values. Since F$_3$ and F$_5$ are 
inverses of F$_4$ and F$_6$, respectively, vertically reversed shapes of the 
former resemble those of the latter. Thus, summing them suppressed their 
transient deviations when deriving the voltage profiles. 
Note, however, that since the tracked values in \name\ depend on the time of 
sampling and its accuracy, the summation did not completely remove the 
deviations, but it sufficed for fingerprinting.

\begin{figure}[!t]
	\begin{subfigure}[!t]{0.49\linewidth}
		\includegraphics[width=\textwidth]{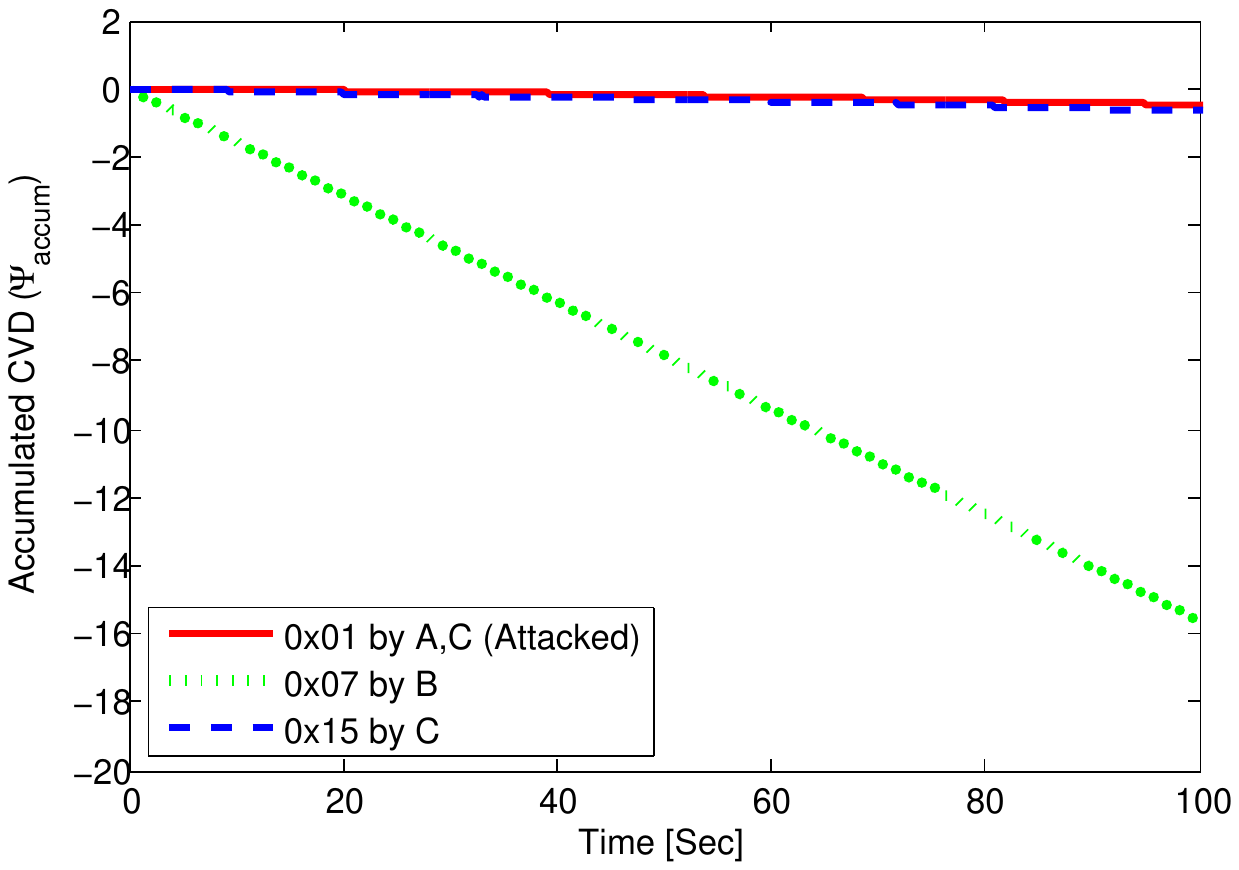}
		\caption{CAN bus prototype.}
		\label{fig:rootcause_fab_proto}
	\end{subfigure}
	~
	\begin{subfigure}[!t]{0.49\linewidth}\centering
		\includegraphics[width=\textwidth]{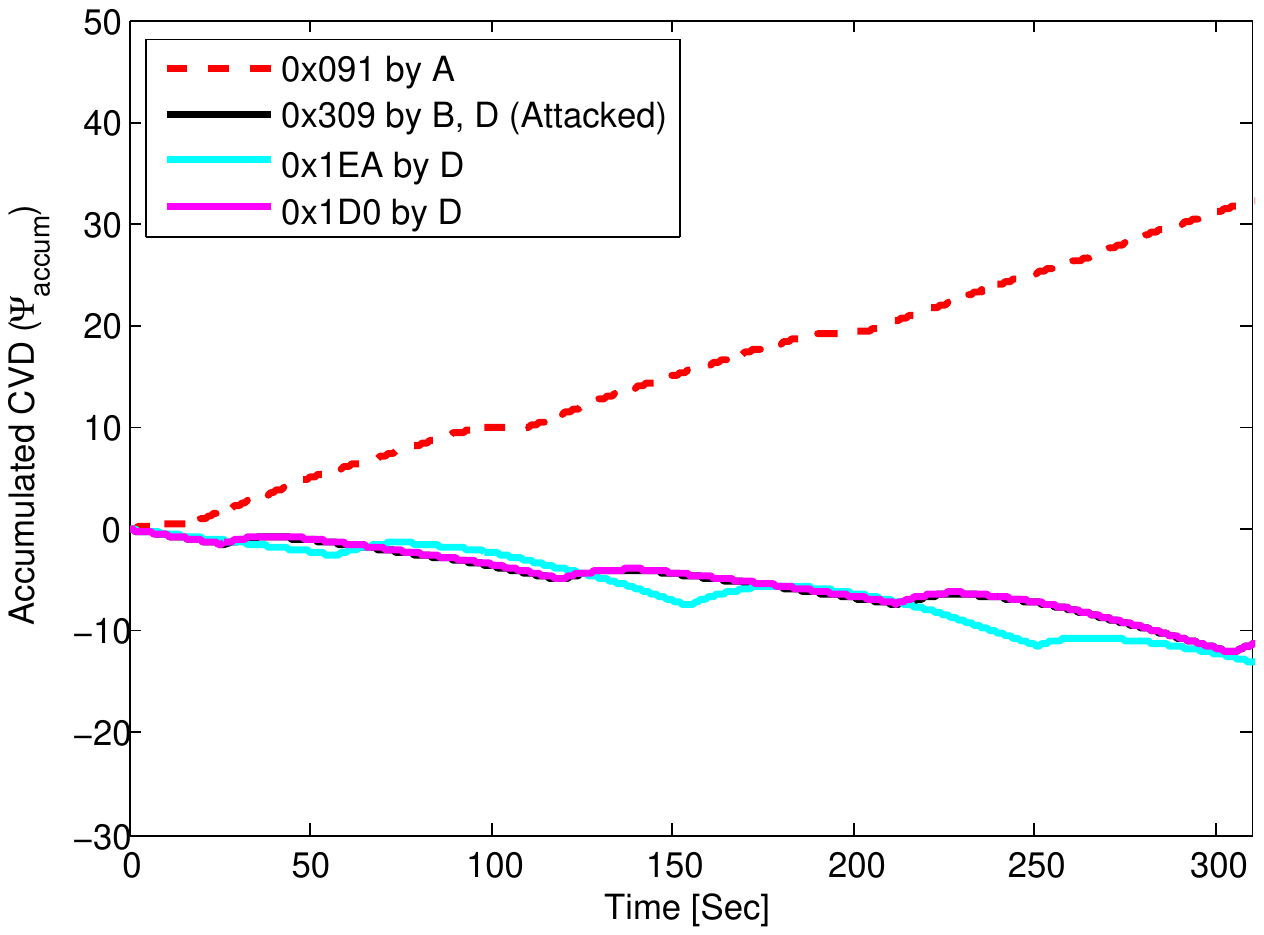}
		\caption{Real vehicle.}
		\label{fig:rootcause_fab_real_att}
	\end{subfigure}
	\caption{\name\ identifying a timing-aware adversary.}
\end{figure}
\subsection{Against a Timing-Aware Adversary}
We evaluated \name's performance of attacker identification in the CAN bus 
prototype and in a real vehicle against a timing-aware adversary. 
We did not evaluate its performance against a naive adversary since 
the timing-aware adversary subsumes his capabilities.

{\bf CAN bus prototype.}
In the CAN bus prototype, we further programmed node $\mathbb{C}$ to be the 
timing-aware adversary who injects not only 0x15 but also attack messages with 
ID=0x01 at a random interval of 10--20ms; injecting messages aperiodically to 
perform arbitrary impersonation and thus evade timing-based 
fingerprinting devices.
Note that 0x01 is also being sent from the legitimate node $\mathbb{A}$ at a 
random interval of 20--200ms. Fig.~\ref{fig:rootcause_fab_proto} shows the 
determined voltage profiles for all three messages during the mounted attack. 
Even though the voltage profile for 0x01 now reflects both the voltage outputs 
from $\mathbb{A}$ and $\mathbb{C}$, since the injection frequency from the 
attacker $\mathbb{C}$ was much higher, the voltage profile for 0x01 changed to 
a profile equivalent to the one shown in 0x15 (sent by $\mathbb{C}$).
As a result, \name\ determined that the transmitters of 0x01 and 0x15 are the 
same, thus identifying the source of the attack to be ECU $\mathbb{C}$.
Note that even when the injection frequency is lower, the attacker ECU can 
be identified by observing the intrusion voltage profile.

{\bf Real vehicle.}
We also evaluated \name's performance against a timing-aware adversary 
in a real vehicle setting. We focus on the results obtained from the 2013 Honda 
Accord for the purpose of more in-depth discussion.
We consider a scenario in which a timing-aware adversary controlling the Honda 
Accord ECU $\mathbb{D}$ attacks ECU $\mathbb{B}$ and also impersonates ECU 
$\mathbb{A}$, i.e., targeted impersonation. 
Thus, from the vehicle, \name\ acquired voltage instances and profiles of the 
monitored messages: 0x091 sent from $\mathbb{A}$, 0x309 from $\mathbb{B}$, and 
\{0x1EA, 0x1D0\} from $\mathbb{D}$.
To generate the scenario of $\mathbb{D}$ impersonating $\mathbb{A}$ (while 
attacking $\mathbb{B}$), $\mathbb{V}$ was further programmed to record only
every 4-th message of 0x091 (sent by $\mathbb{A}$ every 15ms), and every 3rd 
message of 0x1D0 (sent by $\mathbb{D}$ every 20ms) as its ID to be 0x309. 
This was to emulate a scenario where the attacker 
$\mathbb{D}$ injects its attack messages with forged ID=0x309 at a {\em 
	similar} frequency to $\mathbb{A}$, thus attempting to imitate its timing 
behavior for impersonation.

Fig.~\ref{fig:rootcause_fab_real_att} plots the voltage profiles of 
\{0x091, 0x1EA, 0x1D0\} and the intrusion voltage profile of 0x309.
Although the adversary attempted to impersonate ECU $A$, one can see that since 
\name\ fingerprints the transmitter regardless of message timings, the 
intrusion voltage profile of 0x309 matched the profiles of \{0x1EA, 0x1D0\}. 
As a result, \name\ concluded the attacker to be $\mathbb{D}$.

\subsection{Against a Timing-Voltage-Aware Adversary}
Based on his knowledge of voltage-based fingerprinting devices, a 
timing-voltage-aware adversary could attempt to evade \name\ in two ways.
First, the adversary might perform arbitrary impersonation by 
attacking the vehicle only when voltage-based fingerprints have not been 
updated for a long period of time, i.e., a high model-exam discrepancy.
Next, the adversary might also perform targeted impersonation by changing its 
voltage output levels so as to imitate some specific ECUs' voltage output 
behavior.

\subsubsection{Arbitrary impersonation}\label{sec:adjusteval}

In most cases of a timing-voltage-aware adversary performing arbitrary 
impersonation,  \name\ accordingly/adaptively updates the voltage 
profiles and can thus correctly identify the attacker. One corner case, 
however, in detecting the adversary would be when he performs arbitrary 
impersonation by attacking the vehicle only after a long idle period. 
To verify \name's reaction to such an adversary, we evaluated the following 
scenario.
We first obtained the voltage profiles of 0x091 and 0x191 from the 2013 Honda 
Accord while driving the vehicle for approximately 10 mins. 
After 8 and 10 days had elapsed, we again obtained their profiles; the 
average temperatures during the three days were 14.4$\celsius$, 7.7$\celsius$, 
and 12.2$\celsius$, respectively. In between the three update dates, the 
vehicle was driven 700 miles and 40 miles to generate (on purpose) the 
considered scenario where the voltage profiles might be outdated, thus becoming
a chance for the timing-voltage-aware adversary to perform arbitrary 
impersonation.

\begin{figure}[!t]
	\centering
	\begin{subfigure}[h]{0.49\linewidth}
		\centering
		\includegraphics[width=\textwidth]{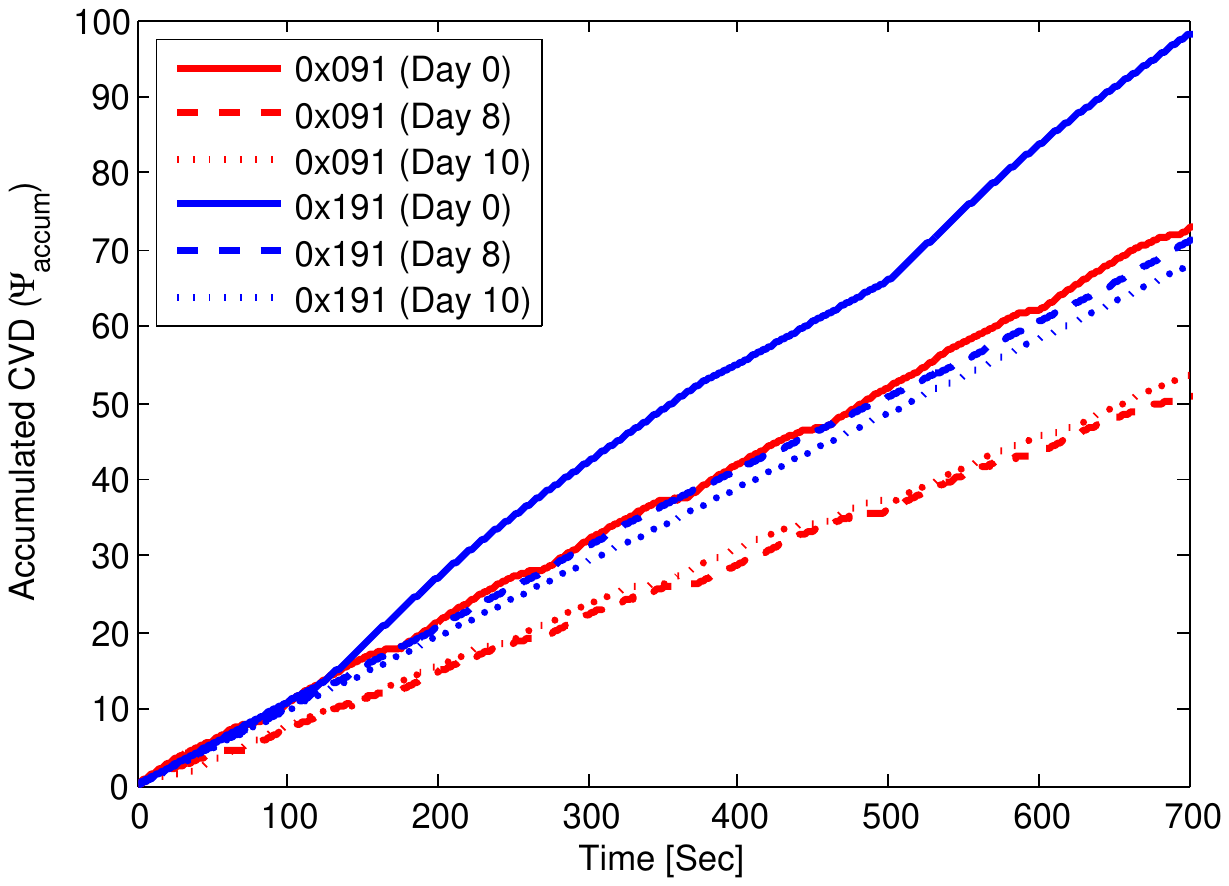}
		\caption{Before adjustment.}
		\label{fig:nonadjust}
	\end{subfigure}%
	~
	\begin{subfigure}[h]{0.49\linewidth}
		\centering
		\includegraphics[width=\textwidth]{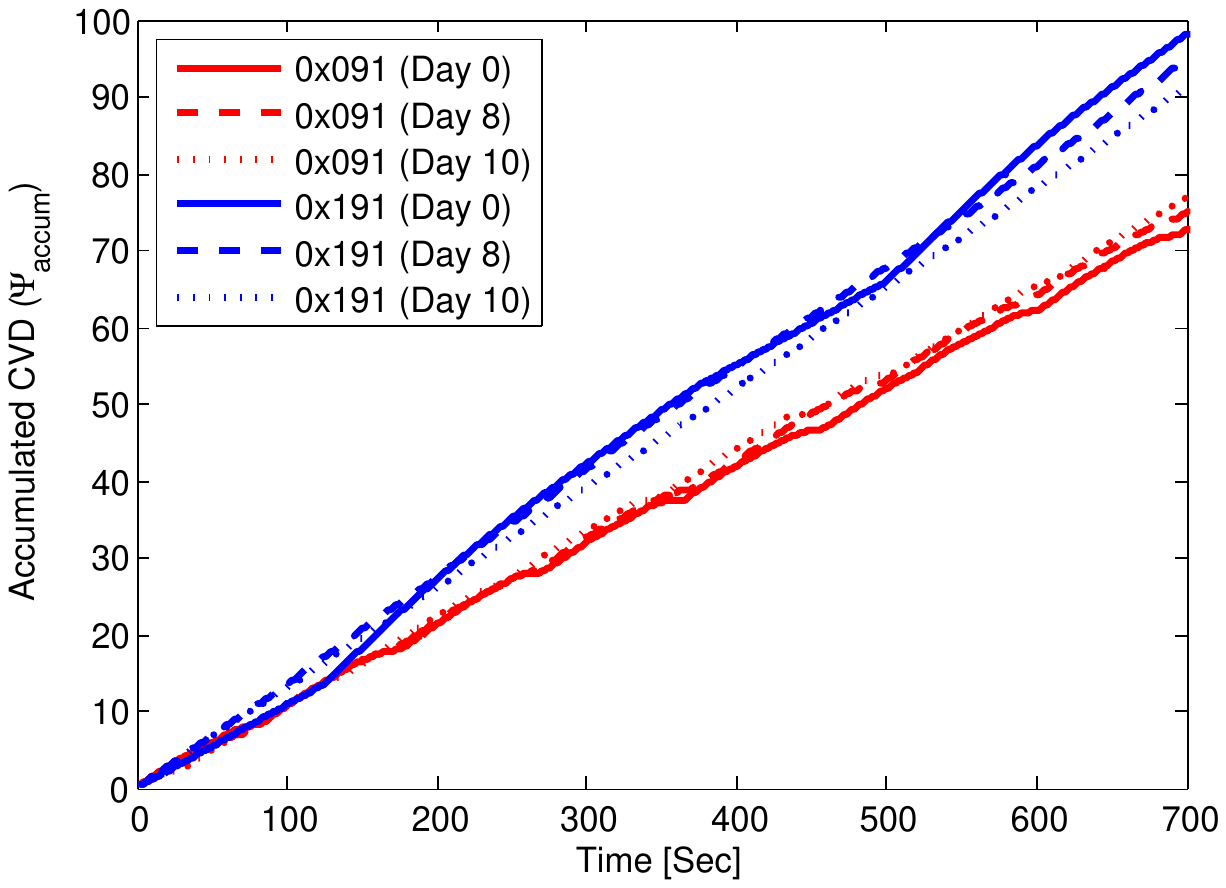}
		\caption{After adjustment.}
		\label{fig:adjust}
	\end{subfigure}
	\caption{Adjusting voltage profiles of \{0x091, 0x191\}.}
	\label{fig:adjust_all}
\end{figure}

Fig.~\ref{fig:nonadjust} shows the acquired voltage profiles corresponding to 
messages 0x091 and 0x191 on the three different dates. 
The initial profiles obtained were found different from those 
obtained on the 8-th and 10-th elapsed days, whereas the latter two were 
equivalent. One interesting observation, however, was that the voltage 
profiles of both message IDs were decreased by the {\em same} amount. 
The changes we observed were due to a slight shift in all ECUs' $V_{CC}$ 
--- most probably due to the change in the battery state after the long 700 
miles driving. In such a case, as we discussed in Section~\ref{sec:adjust}, 
\name\ adjusts its voltage profiles. Once such an adjustment was made, we 
obtained the results shown in Fig.~\ref{fig:adjust}, where all voltage profiles 
were properly aligned. 
This result shows that \name\ is capable of handling cases where a 
timing-voltage-aware adversary performs arbitrary impersonation right after a 
vehicle's long idle period.

\begin{figure}[!t]
	\centering
	\begin{subfigure}[t]{0.48\linewidth}
		\centering
		\includegraphics[width=\textwidth]{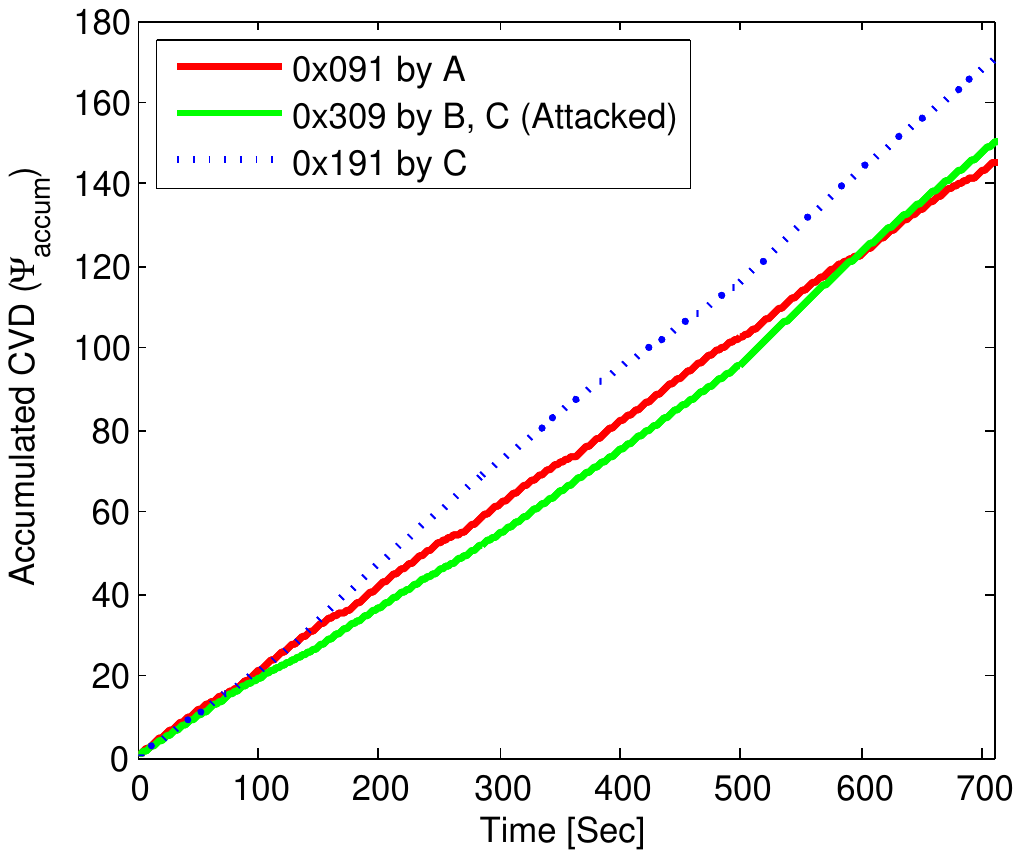}
		\caption{Targeted impersonation.}
		\label{fig:verf1}
	\end{subfigure}%
	~
	\begin{subfigure}[t]{0.5\linewidth}
		\centering
		\includegraphics[width=\textwidth]{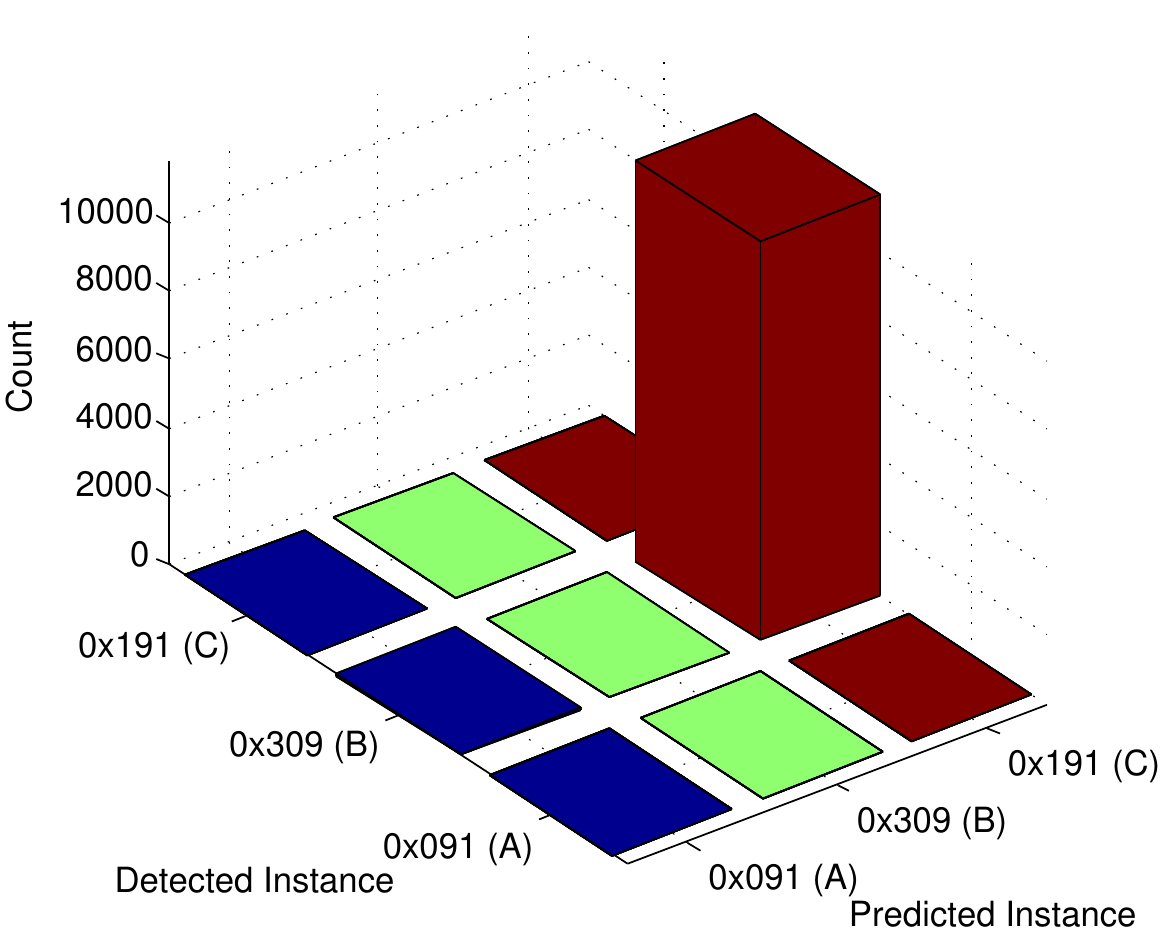}
		\caption{Predicting the attacker ECU.}
		\label{fig:verf2}
	\end{subfigure}
	\caption{Efficacy of \name's Phase 4 execution.}
\end{figure}

\subsubsection{Targeted impersonation}
Under scenarios where a timing-aware adversary performs arbitrary/targeted 
impersonation or where a timing-voltage-aware adversary performs an arbitrary 
impersonation, \name\ can correctly identify him solely based on voltage 
profiles, i.e., within Phase 3. 
It could be much more challenging for \name\ (requiring to run Phase 4) when 
a timing-voltage-aware adversary performs a {\em targeted} 
impersonation, i.e., trying to imitate some specific ECU's voltage outputs.
Since the adversary creates a situation of at least two ECUs having similar 
voltage profiles (i.e., not unique), targeted impersonation would be 
more difficult for \name\ to handle than arbitrary impersonation.
We evaluated how \name\ performs against such an adversary via (1) vehicle 
experiments and (2) simulations based on vehicle data.

{\bf Experiment-based evaluation.}
In the real vehicle setting, to generate a case which reflects 
a timing-voltage-aware adversary performing targeted impersonation, we 
considered the following scenario in the Honda Accord: adversary's ECU 
$\mathbb{C}$, which usually sends 0x191, injects attack messages with ID=0x309, 
thus attacking its original sender $\mathbb{B}$ and at the same time imitating 
$\mathbb{A}$'s voltage output behavior.
In generating such a scenario, evaluation settings were similar to the 
previous ones, except that \name\ recorded every $10n$-th
message of 0x191 as its ID to be 0x309. This was to generate the 
voltage profile of 0x309 to be similar to $\mathbb{A}$'s as in 
Fig.~\ref{fig:verf1}; $\mathbb{C}$ impersonates $\mathbb{A}$. 
To introduce ambiguity in the decision, we do not use the 
intrusion voltage profile in this evaluation.
In such a case, if only voltage profiles are exploited, \name\ 
might consider ECU $\mathbb{A}$ to be the attacker.
However, \name\ deals with this in Phase 4 by 
using voltage instances as machine classifier's input. 
In this evaluation, we used a 200-tree Random Forest classifier with 50\% of 
the acquired data until detecting the intrusion as its training set.

Fig.~\ref{fig:verf2} shows the number of {\em misclassified} voltage instances 
by the Random Forest classifier. It shows that \name\ misclassified 
a large number of 0x309's voltage instances as those of 0x191.
That is, even when the voltage profiles of 0x091 and 0x309 were similar, since 
\name\ observed the measurements in a momentary manner and the adversary was 
incapable of precisely matching them, the attack source was correctly 
identified as $\mathbb{C}$, i.e., transmitter of 0x191, not $\mathbb{A}$.
This validates that by using voltage instances as machine 
classifiers' inputs, \name\ can prevent targeted impersonation by a 
timing-voltage-aware adversary. 

By the Birthday paradox, at least two ECUs may {\em naturally} have
similar voltage profiles. However, since \name\ was feasible to distinguish 
them via machine classifiers, profile collision can be mitigated.

\begin{table}[!t]
	\centering
	\scriptsize
	\begin{tabular}{c*{6}{|E}|} 
		\multicolumn{1}{c}{} & 
		\multicolumn{1}{c}{$\mathbb{A}$} & 
		\multicolumn{1}{c}{$\mathbb{B}$} & 
		\multicolumn{1}{c}{$\mathbb{C}$} & 
		\multicolumn{1}{c}{$\mathbb{D}$} &
		\multicolumn{1}{c}{$\mathbb{E}$} & 
		\multicolumn{1}{c}{$\mathbb{F}$} \\ 
		\hhline{~*6{|-}|}
		$\mathbb{A}$ & 100 & 0 & 0 & 0 & 0 & 0 \\ \hhline{~*6{|-}|}
		$\mathbb{B}$ & 0 & 99.3 & 0 & 0 & 0.7 & 0\\ \hhline{~*6{|-}|} 
		$\mathbb{C}$ & 0 & 0 & 100 & 0 & 0 & 0 \\ \hhline{~*6{|-}|}
		$\mathbb{D}$ & 0 & 0 & 0 & 100 & 0 & 0\\ \hhline{~*6{|-}|}
		$\mathbb{E}$ & 0 & 0 & 0 & 0 & 100 & 0 \\ \hhline{~*6{|-}|}
		$\mathbb{F}$ & 0 & 0 & 0 & 0 & 0 & 100 \\ \hhline{~*6{|-}|}
	\end{tabular}
	\caption*{(a) "Honda Accord" attack dataset.}
	\vspace{0.5cm}
	\begin{tabular}{c*{11}{|E}|} 
		\multicolumn{1}{c}{} & 
		\multicolumn{1}{c}{$\mathbb{A}$} & 
		\multicolumn{1}{c}{$\mathbb{B}$} & 
		\multicolumn{1}{c}{$\mathbb{C}$} & 
		\multicolumn{1}{c}{$\mathbb{D}$} &
		\multicolumn{1}{c}{$\mathbb{E}$} & 
		\multicolumn{1}{c}{$\mathbb{F}$} &
		\multicolumn{1}{c}{$\mathbb{G}$} &
		\multicolumn{1}{c}{$\mathbb{H}$} &
		\multicolumn{1}{c}{$\mathbb{I}$} &
		\multicolumn{1}{c}{$\mathbb{J}$} &
		\multicolumn{1}{c}{$\mathbb{K}$} \\		\hhline{~*9{|-}*2{|-}|}
		$\mathbb{A}$ & 100 & 0 & 0 & 0 & 0 & 0 & 0 & 0 & 0 & 0 & 0\\ 
		\hhline{~*9{|-}*2{|-}|}
		$\mathbb{B}$ & 1.5 & 98.5 & 0 & 0 & 0 & 0 & 0 & 0 & 0 & 0 & 0\\ 
		\hhline{~*9{|-}*2{|-}|}
		$\mathbb{C}$ & 0 & 0 & 100 & 0 & 0 & 0 & 0 & 0 & 0 & 0 & 0\\ 
		\hhline{~*9{|-}*2{|-}|}
		$\mathbb{D}$ & 0 & 0 & 0 & 100 & 0 & 0 & 0 & 0 & 0 & 0 & 0\\
		\hhline{~*9{|-}*2{|-}|}
		$\mathbb{E}$ & 0 & 0 & 0 & 0 & 100 & 0 & 0 & 0 & 0 & 0 & 0\\ 
		\hhline{~*9{|-}*2{|-}|}
		$\mathbb{F}$ & 0 & 0 & 0 & 0 & 0 & 100 & 0 & 0 & 0 & 0 & 0 \\ 
		\hhline{~*9{|-}*2{|-}|}
		$\mathbb{G}$ & 0 & 0 & 0 & 0 & 0 & 0 & 100 & 0 & 0 & 0 & 0 \\ 
		\hhline{~*9{|-}*2{|-}|}
		$\mathbb{H}$ & 0 & 0 & 0 & 0 & 0 & 0 & 0 & 100 & 0 & 0 & 0 \\ 
		\hhline{~*9{|-}*2{|-}|}
		$\mathbb{I}$ & 0 & 0 & 0 & 0 & 0 & 0 & 0 & 0 & 100 & 0 & 0 \\ 
		\hhline{~*9{|-}*2{|-}|}
		$\mathbb{J}$ & 0 & 0 & 0 & 0 & 0 & 0 & 0 & 0 & 0 & 100 & 0 \\ 
		\hhline{~*9{|-}*2{|-}|}
		$\mathbb{K}$ & 0 & 0 & 0 & 0 & 0 & 0 & 0 & 0 & 3.2 & 0 & 96.8 \\ 
		\hhline{~*9{|-}*2{|-}|}
	\end{tabular}
	\caption*{(b) "Honda Accord + Chevrolet Trax" attack dataset.}
	\caption{Confusion matrix of \name\ [Unit: \%].}
	\label{tab:confuse}
\end{table}

{\bf Simulation-based evaluation.}
In addition to the scenario shown in Fig.~\ref{fig:verf1}, which we 
evaluated via real vehicle experiments, there could be different ways in which 
a timing-voltage-aware adversary might perform targeted impersonation.
For example, the adversary might heat up or cool down his ECU to match some 
other ECUs' voltage profiles, even before he starts injecting attack messages.
Thus, we conducted a more in-depth evaluation as follows. 
Based on the 35-min data of voltage instances output by the 
Honda Accord's 6 ECUs and those output by the Chevrolet Trax's 5 ECUs, two 
attack datasets were constructed to each contain 1000 different ``targeted 
impersonation'' attempts by a timing-voltage-aware adversary.
We refer to Honda Accord's ECUs as $\mathbb{A}$--$\mathbb{F}$ and Chevrolet 
Trax's ECUs as $\mathbb{G}$--$\mathbb{K}$.
The first dataset was based on only voltage instances of 
$\mathbb{A}$--$\mathbb{F}$ whereas the second was based on data from both 
vehicles, assuming that $\mathbb{A}$--$\mathbb{K}$ lie in the same vehicle.
Such an assumption was made to evaluate how \name\ performs when the number of 
ECUs increases.
Each impersonation attempt was constructed by (1) randomly choosing one ECU to 
be the adversary and another to be the victim, then (2) randomly choosing the 
times when the adversary starts to change his voltage outputs and (later) when 
to start attacking the victim, and finally (3) steadily shifting the 
adversary's voltage instance values (when it starts impersonation) so that 
his voltage profile matches the victim's, i.e., profile collision, before 
mounting an attack. 
Note, however, that such a shift does not make their instantaneous instances to 
be equivalent. As we discussed in Section~\ref{sec:attackmodel}, although an 
adversary may match the target's profile, it would be very difficult for him to 
precisely follow the target's instantaneous behaviors (e.g., transient changes 
due to temperature). This way, we were able to 
emulate a scenario where the adversary first imitates some specific ECU's 
voltage output behavior and then injects attack messages.

Table~\ref{tab:confuse}a shows the confusion matrix of \name\ when 
identifying the attacker of the 1000 targeted impersonation attempts in the 
first attack dataset. For identification, \name\ not only used voltage profiles 
but also a 200-tree Random Forest with voltage instances as its input. 
Again, half of the data until an attack was detected was used as the training 
set. Thanks to \name's analysis of the adversary's 
impersonation attempts from two different viewpoints --- ECU's usual voltage 
output behavior via voltage profiles and its momentary behavior via voltage 
instances --- \name\ was able to identify the attacker with only a 
0.2\% false identification rate.
Even when \name\ was evaluated based on our second attack dataset, which had 
11 ECUs, \name\ identified the attacker with a 0.3\% false identification rate 
where the confusion matrix is shown in Table~\ref{tab:confuse}b.
Albeit the increased number of ECUs, \name's false identification rate 
increased only by 0.1\%, thus corroborating its effectiveness.
Note that such false rates reflect \name's capability and robustness against 
the most skillful adversary who is aware of timing and voltage, i.e., 
the timing-voltage-aware adversary. 
Thus, \name's false rate against {\it all} types of the considered 
adversaries --- including the naive and timing-aware adversaries ---
would be much lower.

One can also interpret such good performance of \name\ equivalent to its 
effectiveness in mitigating (naturally occurred) profile collision.


\section{Discussion}\label{sec:disc}
{\bf Number of ECUs on CAN.}
As of 2017, the average vehicle is reported to have approximately 25 
ECUs, while luxury cars have approximately 50~\cite{nxpreport}, but 
{\em not all} of them on CAN; some are installed on LIN, MOST, etc.
Moreover, to accommodate a large (increasing) number of ECUs on
bandwidth-limited CAN, each vehicle is equipped with {\em multiple} CAN 
buses~\cite{fostercan}.
Accordingly, network architectures of various modern vehicles (Audi 
A8, Honda Accord, Jeep Cherokee, Infiniti Q50, etc.) are shown to have 
3$\sim$20 ECUs {\em per} CAN bus~\cite{miller2}; a similar figure to which 
we considered in our evaluations. 
Hence, if \name\ was installed on each CAN bus in a vehicle, profile 
collision within that bus is much less likely to occur than the case when 
all ECUs are (considered to be) installed on one single CAN bus. 
Even in such a case with profile collisions, \name\ can still handle it via 
the execution of its Phase 4.

{\bf Multiple ECUs per ID.}
\name\ may underperform when multiple ECUs are assigned to send messages 
with the same ID, albeit unusual/rare. 
For example, although message ID=0x040 is 
scheduled to be sent, in turn, by ECUs $\mathbb{A}$--$\mathbb{D}$, 
\name\ would construct only one voltage profile for 0x040. 
However, if such scheduling information is known in advance
(e.g., every $4n$-th message of 0x040 is sent by $\mathbb{D}$), which is in 
fact defined by the car-makers, then \name\ could construct voltage profiles 
accordingly, thus solving the problem.

{\bf Intrusion Detection.}
Timing-based IDSs exploit the {\em periodic} nature of CAN messages and 
thus suffice to detect attacks on periodic messages, but fail to detect 
attacks on {\em aperiodic} ones. Since \name\ determines the transmitter ECU 
based on voltages, similarly to \cite{sourceiden,voltarxiv}, it can 
complement those IDSs in detecting intrusions. 
However, since most in-vehicle messages are 
periodic~\cite{cids} and thus most intrusions are detectable,  
\name's potential is maximized when it is used for attacker identification.
%

{\bf Attacker from Another In-vehicle Network.}
If the attack originates from a different in-vehicle network  
(e.g., FlexRay, MOST, LIN) inside the vehicle other than CAN, 
the corresponding gateway ECU will be the one that injects attack messages into 
CAN.
\name\ will, therefore, identify that gateway ECU as the attacker, 
since \name\ is designed just for CAN. In such a case, the best  both \name\
and the gateway ECU can do is to look up the message 
routing table (describing which messages/signals to forward to/from), 
and identify the ``compromised network''. 
Handling such a scenario is important in integrating \name\ 
in real vehicles. 

{\bf Limitations.}
For \name\ to identify the attacker ECU, it requires at least one voltage 
profile to use. For the example shown in 
Fig.~\ref{fig:rootcause_fab_real_att}, \name\ referred to the voltage profiles 
of \{0x1EA, 0x1D0\} to determine that the attacker ECU was $\mathbb{D}$. Since 
most ECUs are designated to transmit at least one message ID, one can identify 
the attacker ECU with \name. However, if the compromised ECU does not send any 
messages, \name's attacker identification can be inaccurate. In such a case, 
the best \name\ can do would be obtaining the voltage profile of those ECUs 
during the manufacturing stage and updating them via voltage profile 
adjustments.

\section{Conclusion}\label{sec:conclusion}
State-of-the-art vehicle security solutions lack a key feature of 
identifying the attacker ECU on the in-vehicle network, which
is essential for efficient forensic, isolation, security patching, etc.
To meet this need, we have proposed \name, which
fingerprints ECUs based on voltage measurements. 
Via the ACK learning phase, \name\ obtained correct measurements of voltages 
only from the message transmitters, and exploited them for constructing and 
updating correct voltage profiles/fingerprints.
Using these profiles, we showed via evaluations on a CAN bus prototype and two 
real vehicles that \name\ can identify the attacker ECU with a low false 
identification rate of 0.2\%.
Considering the fact that vehicles are safety-critical, \name\ is an important 
first step toward securing the vehicles and protecting drivers and 
passengers.
\bibliographystyle{IEEEtran}
\bibliography{refer}
\section*{Appendices}
\addcontentsline{toc}{section}{Appendices}
\renewcommand{\thesubsection}{\Alph{subsection}}

\subsection{ACK threshold learning in a real vehicle}\label{sec:acklearn_real}
\name\ first learns the thresholds
which determine whether the measured voltages are from the ACK slot or not,
before outputting voltage instances and profiles.
This was achieved by determining most frequent and
maximum/minimum sets, and exploiting the side lobe which only exists
in the latter. In other words, the existence of such a side lobe (as shown in
Fig.~\ref{fig:acknonack} in a CAN prototype), which represents the
distribution of ACK voltages, is critical in learning the ACK threshold.
Thus, to show that the proposed ACK learning  is feasible even in real vehicles,
through \name, we obtained both most frequent and maximum/minimum sets
for message ID=0x091, which was sent every 10ms by some ECU in the 2013 Honda 
Accord.

\begin{figure}[!h]
	\centering
	\includegraphics[width=0.9\linewidth]{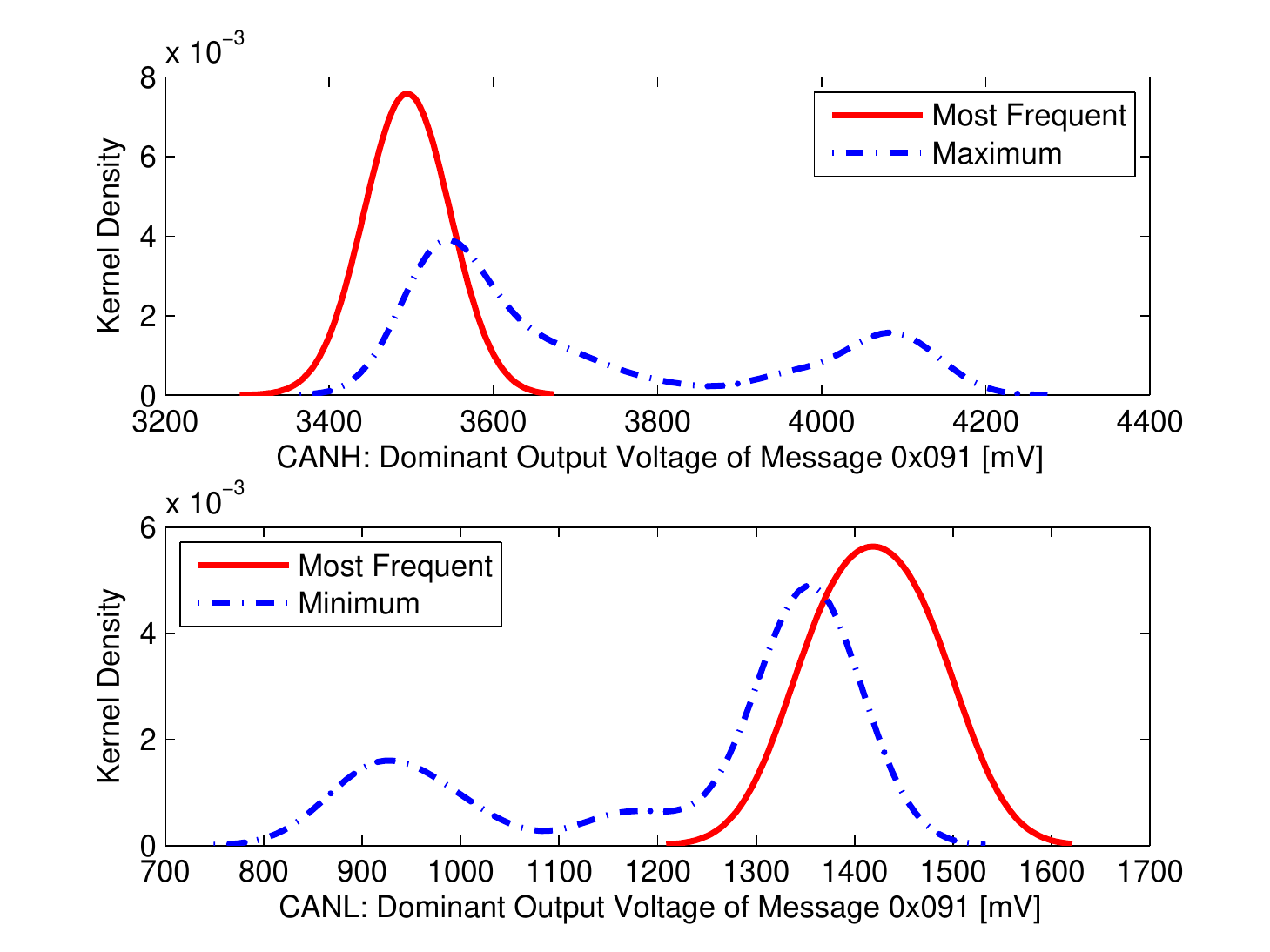}
	\caption{Deriving the ACK thresholds for message 0x091 in the 2013 Honda 
		Accord.}
	\label{fig:acknonack_real}
\end{figure}

Fig.~\ref{fig:acknonack_real} (upper) shows the kernel density plots of the 
most frequent and
maximum sets of the CANH dominant voltages while receiving ID=0x091, and
Fig.~\ref{fig:acknonack_real} (lower) the kernel density of those obtained from 
the CANL line.
One can see that as in the CAN prototype result (Fig.~\ref{fig:acknonack}), 
side lobes exist
in both the CANH and CANL lines. Thus, the proposed ACK learning mechanism in 
\name\ derived the 
refined maximum and minimum sets, $S'_{max}$ and $S'_{min}$, correctly and thus 
derived the 
ACK thresholds of message 0x091 to be 3.844V for CANH outputs and 1.114V for 
CANL outputs.

One interesting observation is how high and low CANH and CANL ACK voltage 
levels are.
In our evaluation of \name\ on the CAN prototype, since we had only 3 nodes 
acknowledging to the message, 
the median of the CANH ACK voltages was 3.514V as shown in 
Fig.~\ref{fig:acknonack}. 
On the other hand, in our experiment on the 2013 Honda Accord, the median of 
the CANH ACK
voltage showed a high voltage level, 4.049V --- much higher than the one 
obtained from the CAN prototype. 
For the CANL ACK voltage, the median was 0.953V. Such a result is due to the 
fact that there were
much more ECUs (compared to 3 in the prototype) inside the vehicle which ACKed 
message 0x091.

\begin{figure}[!t]
	\centering
	\includegraphics[width=0.8\linewidth]{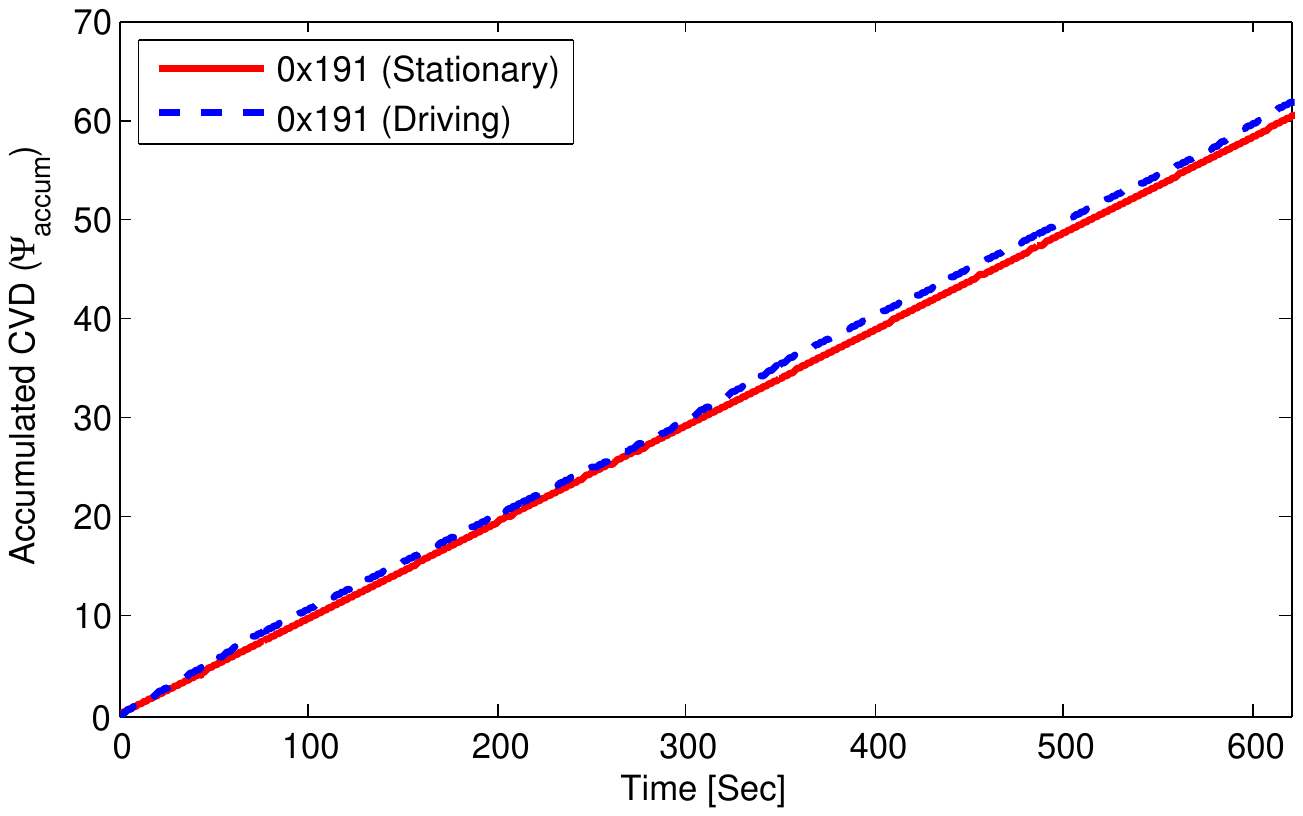}
	\caption{Voltage profiles of message 0x191 when the vehicle was stationary 
		and 
		driven.}
	\label{fig:whiledrive}
\end{figure}

\subsection{Voltage profiles while driving}\label{sec:whiledrive}
We further validate that \name's derived voltage profiles do not depend on 
whether and how the vehicle is driven. 
We first obtained the voltage instances of 0x191 from the 
2013 Honda Accord's CAN bus. At this time of measurement, the vehicle was 
stationary.
Later on that day, instances of 0x191 was once again obtained, but this time 
while driving the vehicle for approximately 10 mins; the same data which we 
used in Section~\ref{sec:adjusteval}.

Fig.~\ref{fig:whiledrive} shows the voltage profiles of 0x191 obtained while
the vehicle was stationary and driven.
One can see that the two voltage profiles are equivalent, even though they were 
measured 
under a different condition.
These are due to the fact that the voltage outputs are much more dependent on 
hardware 
components' characteristics than their momentary conditions.
Moreover, transient deviations incurred from changes in momentary conditions 
would have
been suppressed thanks to how \name\ derives its voltage profiles; summing CVDs 
of CANH and CANL.

\end{document}